\documentstyle[12pt]{article}
\def\hybrid{\topmargin -20pt    \oddsidemargin 0pt
    \headheight 0pt \headsep 0pt
    \textwidth 6.25in   
    \textheight 9.5in   
    \marginparwidth .875in
    \parskip 5pt plus 1pt   \jot = 1.5ex}

\hybrid

\def\baselinestretch{1.2}

\catcode`\@=11

\def\marginnote#1{}
\newcount\hour
\newcount\minute
\newtoks\amorpm
\hour=\time\divide\hour by60
\minute=\time{\multiply\hour by60 \global\advance\minute by-\hour}
\edef\standardtime{{\ifnum\hour<12 \global\amorpm={am}%
    \else\global\amorpm={pm}\advance\hour by-12 \fi
    \ifnum\hour=0 \hour=12 \fi
    \number\hour:\ifnum\minute<10 0\fi\number\minute\the\amorpm}}
\edef\militarytime{\number\hour:\ifnum\minute<10 0\fi\number\minute}

\def\draftlabel#1{{\@bsphack\if@filesw {\let\thepage\relax
   \xdef\@gtempa{\write\@auxout{\string
      \newlabel{#1}{{\@currentlabel}{\thepage}}}}}\@gtempa
   \if@nobreak \ifvmode\nobreak\fi\fi\fi\@esphack}
    \gdef\@eqnlabel{#1}}
\def\@eqnlabel{}
\def\@vacuum{}
\def\draftmarginnote#1{\marginpar{\raggedright\scriptsize\tt#1}}

\def\draft{\oddsidemargin -.5truein
    \def\@oddfoot{\sl preliminary draft \hfil
    \rm\thepage\hfil\sl\today\quad\militarytime}
    \let\@evenfoot\@oddfoot \overfullrule 3pt
    \let\label=\draftlabel
    \let\marginnote=\draftmarginnote
   \def\@eqnnum{(\theequation)\rlap{\kern\marginparsep\tt\@eqnlabel}%
\global\let\@eqnlabel\@vacuum}  }

\def\preprint{\twocolumn\sloppy\flushbottom\parindent 2em
    \leftmargini 2em\leftmarginv .5em\leftmarginvi .5em
    \oddsidemargin -.5in    \evensidemargin -.5in
    \columnsep .4in \footheight 0pt
    \textwidth 10.in    \topmargin  -.4in
    \headheight 12pt \topskip .4in
    \textheight 6.9in \footskip 0pt
    \def\@oddhead{\thepage\hfil\addtocounter{page}{1}\thepage}
    \let\@evenhead\@oddhead \def\@oddfoot{} \def\@evenfoot{} }

\def\numberbysection{\@addtoreset{equation}{section}
    \def\theequation{\thesection.\arabic{equation}}}

\def\underline#1{\relax\ifmmode\@@underline#1\else
    $\@@underline{\hbox{#1}}$\relax\fi}

\def\titlepage{\@restonecolfalse\if@twocolumn\@restonecoltrue\onecolumn
     \else \newpage \fi \thispagestyle{empty}\c@page\z@
    \def\thefootnote{\fnsymbol{footnote}} }

\def\endtitlepage{\if@restonecol\twocolumn \else \newpage \fi
    \def\thefootnote{\arabic{footnote}}
    \setcounter{footnote}{0}}  

\catcode`@=12
\relax

\def\figcap{\section*{Figure Captions\markboth
    {FIGURECAPTIONS}{FIGURECAPTIONS}}\list
    {Figure \arabic{enumi}:\hfill}{\settowidth\labelwidth{Figure 999:}
    \leftmargin\labelwidth
    \advance\leftmargin\labelsep\usecounter{enumi}}}
 \relax
\def\tablecap{\section*{Table Captions\markboth
    {TABLECAPTIONS}{TABLECAPTIONS}}\list
    {Table \arabic{enumi}:\hfill}{\settowidth\labelwidth{Table 999:}
    \leftmargin\labelwidth
    \advance\leftmargin\labelsep\usecounter{enumi}}}
 \relax
\def\reflist{\section*{References\markboth
    {REFLIST}{REFLIST}}\list
    {[\arabic{enumi}]\hfill}{\settowidth\labelwidth{[999]}
    \leftmargin\labelwidth
    \advance\leftmargin\labelsep\usecounter{enumi}}}
 \relax
\makeatletter
\newcounter{pubctr}
\def\publist{\@ifnextchar[{\@publist}{\@@publist}}
\def\@publist[#1]{\list
    {[\arabic{pubctr}]\hfill}{\settowidth\labelwidth{[999]}
    \leftmargin\labelwidth
    \advance\leftmargin\labelsep
    \@nmbrlisttrue\def\@listctr{pubctr}
    \setcounter{pubctr}{#1}\addtocounter{pubctr}{-1}}}
\def\@@publist{\list
    {[\arabic{pubctr}]\hfill}{\settowidth\labelwidth{[999]}
    \leftmargin\labelwidth
    \advance\leftmargin\labelsep
    \@nmbrlisttrue\def\@listctr{pubctr}}}
 \relax
\makeatother
\newskip\humongous \humongous=0pt plus 1000pt minus 1000pt
\def\caja{\mathsurround=0pt}
\def\eqalign#1{\,\vcenter{\openup1\jot \caja
    \ialign{\strut \hfil$\displaystyle{##}$&$
    \displaystyle{{}##}$\hfil\crcr#1\crcr}}\,}
\newif\ifdtup

\relax

\def\thefootnote{\fnsymbol{footnote}}
\def\be{\begin{equation}}
\def\ee{\end{equation}}
\def\ba{\begin{eqnarray}}
\def\ea{\end{eqnarray}}
\def\d{\partial}
\def\dd{{\rm d}}

\def\s{\sigma}

\def\square{\hbox{{$\sqcup$}\llap{$\sqcap$}}}   

\def\P{\Phi}

\def\ub{{\bar u}}
\begin{document}
\renewcommand{\theequation}{\thesection.\arabic{equation}}
\begin{titlepage}
\begin{center}

\hfill CERN-TH.6975/93\\
\hfill HUB-IEP-93/3\\
\hfill LPTENS 93/31\\
\hfill hep-th/9308124\\
\vskip .1in

{\large \bf A Large Class of New
Gravitational and Axionic Backgrounds for
Four-dimensional Superstrings}
\vskip .5in

{\bf E. Kiritsis, C. Kounnas\footnote{On leave from Ecole Normal
Superieure,
24, rue Lhomond, 75231 Paris Cedex 05, France.} }
\vskip .1in
{\em CERN, Geneva, SWITZERLAND}
\vskip .1in
and
\vskip .1in
{\bf D. L\"ust}

\vskip .1in
{\em Humboldt Universit\"at zu Berlin\\
Fachbereich Physik\\
D-10099 Berlin, GERMANY}

\end{center}

\vskip .7in

\begin{center} {\bf ABSTRACT } \end{center}
\begin{quotation}\noindent
A large class of new 4-D superstring vacua
with non-trivial/singular geometries, spacetime supersymmetry
and other background fields (axion, dilaton) are found.
Killing symmetries are generic and are associated with
non-trivial dilaton and
antisymmetric
tensor fields. Duality symmetries preserving N=2 superconformal
invariance
are employed to generate a large class of explicit metrics for
non-compact
4-D Calabi-Yau manifolds with Killing symmetries.
We comment on some of our solutions which have interesting
singularity
properties and cosmological interpretation.
\end{quotation}
\vskip1.0cm
CERN-TH.6975/93 \\
August 1993\\
\end{titlepage}
\vfill
\eject
\def\baselinestretch{1.2}
\baselineskip 16 pt
\noindent
\section{Introduction}
\setcounter{equation}{0}

String theory is a promising candidate for a theory unifying all
the fundamental interactions, including quantum gravity.
Gauge and diffeomorphism symmetries appear naturally and the theory
has very smooth ultraviolet behavior.
However, the theory is still conceptually incomplete, in the sense
that,
a background independent string field theory, where the full string
symmetry
is manifest, is not at hand.
In particular, we have no way, so far, to understand non-perturbative
effects (except in some toy bosonic models in 2-D, \cite{2d}).
The only way we know, up to now, to deal with the theory, is
in the first
quantized perturbative framework, where one starts with a classical
solution to the theory, which is given by any CFT with the
appropriate
central charge. One then calculates in string perturbation theory
the S-matrix of fluctuations around that classical solution.
In particular, it is known that, unless the spacetime physics is at
least
N= supersymmetric, then the initial perturbative vacuum is in general
unstable and therefore perturbation theory around it is
ill defined.
Thus, in perturbation theory, only classical solutions with at least
N=
spacetime supersymmetry are guaranteed to be stable.

There are some extra problems in order to connect superstring theory
predictions
at low energy to the observed world.
The first is that there is always a scalar excitation
in the theory, the dilaton, which is massless to all orders
in perturbation theory. There are good reasons to believe that such a
massless
scalar does not exists in our world.
The second is that, at low energy, there is no supersymmetry in the
observed
spectrum of particles.
In fact, the breaking of space-time supersymmetry in string theory
was extensively discussed in the past,
where essentially two kinds of mechanisms were considered:
the ``perturbative'' breaking via the Scherk-Schwarz
mechanism \cite{SSc} and non-perturbative supersymmetry breaking by
gaugino condensation at the level of the string effective field
theory
\cite{gaugi}.
However, unless we have a good string field theory which is
background independent, there is little chance that we will have a
{\it fundamental} understanding of the resolution of the two previous
problems.

However, there are still some important conceptual questions one can
ask in string theory.
So far, in field theory there is not even a consistent perturbative
framework, in order to address questions concerning quantum gravity.
On the other hand, string theory provides also a theory of quantum
gravity
without ultraviolet problems. Even in the present limited
formulations of string theory, there is still room to address
questions which have perplexed
physicists in the past decades concerning in particular issues like
set-up of scattering theory in curved spacetimes, black holes,
Hawking
radiation etc.

Several efforts focused towards this direction in the recent past.
One expects that the physics of gravitation at short distances,
as described by string theory, would be radically different from that
at large distances and scales. The probes here are entities (strings)
with non-zero
size, and there are quantitative indications that the concept of
classical metric is not good enough to describe even qualitatively
the physics at
regions of spacetime which are strongly curved \cite{K2,C1}.
Such indications come from the existence of duality symmetries in
many
non-trivial classical solutions of string theory (where the metric,
dilaton
and other background fields are non-trivial functions of
coordinates).
These symmetries imply that the background fields by themselves do
not
uniquely describe the ground state, but stringy probes, excited in
different
states, ``see" different background fields.

In this work, we will address a specific aspect of these problems,
namely, we will try to analyse as exhaustively as possible, 4-d
string solutions
which have the following properties: The metric and other background
fields
are non-trivial, and the solutions are perturbatively stable,
equivalently, they have spacetime supersymmetry.
In the context of the type II superstring, this translates to the
statement
that the worldsheet CFT has (2,2) superconformal invariance (in which
case
the spacetime theory in 4-D has N=2 supersymmetry) \cite{BD,G}.
Via the heterotic string map \cite{het,G},
such type II solutions generate a class of heterotic
string models with $N=1$ space time supersymmetry.

During recent years, several efforts were focused
on $N=2$ superconformal backgrounds in the context
of (heterotic) string
compactifications leading to strings with flat four-dimensional
space time \cite{4d,het,G}.
In these works the discussion mainly focussed
on compact spaces without torsion and with constant
dilaton field, i.e. tori, orbifolds, the $K_3$
manifold and numerous Calabi-Yau spaces.

In this paper we will provide a relatively systematic discussion on
supersymmetric string backgrounds with $N=2$ or $N=4$ superconformal
symmetry,
based on compact as well as
non-compact spaces plus
non-trivial antisymmetric tensor-field
and non-constant dilaton. Thus we will extend in a more systematic
way
the exact N=4 solution constructed recently, \cite{Ca,C2}.
In the case of vanishing antisymmetric
tensor fields the metric is known to describe a compact/non-compact
K\"ahler
space \cite{kahler}.
In order to introduce non-constant antisymmetric
tensor fields,  we have to use a 2-d superspace action,
which includes chiral as well as twisted chiral
superfields \cite{GHR}.
In contrast to the compact Calabi-Yau spaces, almost all backgrounds
with non-trivial dilaton field will possess Killing symmetries.
Our main focus will be such backgrounds with isometries which turn
out to be
generic in 4-D (remember that in 2-D all string backgrounds have
killing symmetries).
In addition,
most of such backgrounds exhibit singularities on some hypersurface
in spacetime.
These spaces, which can be regarded as
generalizations of the two-dimensional black-hole considered
in \cite{bh}, provide gravitational instantons, black-hole type
backgrounds, worm-holes and cosmological solutions in four
and higher dimensions \cite{bh1,Ca,rey,C1,C2,DVV}.

A key to the proper understanding of string propagation
on curved spaces is provided by duality symmetries.
As mentioned above, duality symmetries relate backgrounds which are
geometrically or even topologically different \cite{GK} but
nevertheless
correspond to the same superconformal field theory (there is a
non-local transformation in the $\sigma$-model variables that maps
the $\sigma$-model
to its dual \cite{K2}).
For example, a singular space can be dual to non-singular space,
as first shown for the Euclidean two-dimensional
black-hole \cite{Gi,DVV,K1}. Although it is not known in general,
what are
all dual equivalent backgrounds starting from a given background,
the construction of the dual spaces is straightforward
if the original space possesses some isometries \cite{BU,odd,GR}
(although this is not necessary, \cite{K2,K1}).
For the cases with extended world-sheet supersymmetries
it was shown \cite{LR,GHR} that the duality symmetry originating from
a $U(1)$
isometry can be understood as replacing chiral (twisted chiral)
superfields by twisted chiral (chiral) superfields by a
Legendre transformation. This duality transformation preserves the
$N=2$ superconformal invariance of the theory.
Some interesting
examples of dual backgrounds were already constructed this way
\cite{RSS}.

The structure of this paper is as follows.
In the next chapter we will set up the $N=2$ ($N=4$) supersymmetric
$\s$-model using chiral and twisted chiral superfields.
We end this section
with the construction the dual spaces assuming that the original
space has some $U(1)$ isometries.
Section three deals with the K\"ahlerian solutions of
the string field equations. For non-trivial dilaton field
the spaces are in general non Ricci-flat and possess Killing
symmetries (isometries).
We also construct the corresponding dual spaces which are in
general non-K\"ahlerian.
In section four we consider solutions which from the
beginning possess torsion and are therefore non-K\"ahlerian.
Performing a duality transformation we obtain additional new
classes of K\"ahlerian spaces where many of them are Ricci-flat.
Section five discusses some aspects of S duality \cite{S}
pertinent to some of our solutions.
Finally section six contains our conclusions and further remarks.
In the appendix we discuss some constraints for the superpotential
which are of direct interest to N=2 LG models \cite{Le}, with
non-trivial
metrics.

During the paper, we will mainly
focus on four-dimensional spaces which appear to have most
physical relevance.

\renewcommand{\thesubsection}{\thesection.\arabic{subsection}}
\renewcommand{\theequation}{\thesubsection.\arabic{equation}}

\section{The $N=2$ ($N=4$) Background and $U(1)$ Duality
Transformations}
\setcounter{subsection}{0}
\subsection{The $N=2$ supersymmetric 2-d action}
\setcounter{equation}{0}
The most general $N=2$ superspace action
for $m$ chiral superfields $U_i$ ($i=1,\dots , m$) and
$n$ twisted chiral superfields $V_p$ ($p=1,\dots ,n$)
in two dimensions is determined by a single real function
$K(U_i,\bar U_i,V_p,\bar V_p)$:
\be
S={1\over 2\pi \alpha '}\int{\rm d}^2xD_+D_-\bar D_+\bar D_-
K(U_i,\bar U_i,V_p,\bar V_p).\label{action}
\ee
The fields $U_i$ and $V_p$ obey a chiral or twisted chiral
constraint
\be
\bar D_\pm U_i=0,\qquad\bar D_+V_p=D_-V_p=0.\label{constraint1}
\ee
The action (\ref{action}) is invariant, up to total derivatives,
under quasi-K\"ahler gauge transformations
\be
K\rightarrow K+f(U_i,V_p)+g(U_i,\bar V_p)
+\bar f(\bar U_i,\bar V_p)+\bar g(\bar U_i,V_p).\label{inva}
\ee

To see the background interpretation of the theory it is convenient
to write down the purely bosonic part of the superspace action
(\ref{action}):
\be
\eqalign{
S=-{1\over 2\pi\alpha '
}\int{\rm d}^2x&\lbrack K_{u_i\bar u_j}\partial^a u_i
\partial_a\bar u_j-K_{v_p\bar v_q}\partial^a v_p\partial_a\bar v_{ q}
\cr& +\epsilon_{ab}(K_{u_i\bar v_p}\partial_a u_i\partial_b\bar v_{
p}
+K_{v_p\bar u_i}\partial_a v_p\partial_b\bar u_{i})\rbrack,\cr}
\label{bosonic}
\ee
where
\be
\eqalign{
&K_{u_i\bar u_j}={\partial^2K\over\partial U_i\partial\bar
U_j},\qquad
K_{v_p\bar v_q}={\partial^2K\over\partial V_p\partial\bar V_q},\cr &
K_{u_i\bar v_p}={\partial^2K\over\partial U_i\partial\bar V_p},\qquad
K_{v_p\bar u_i}={\partial^2K\over\partial V_p\partial\bar
U_i}.\cr}\label{
metrics}
\ee
Here $u_{i}$ is the lowest component of the superfield $U_{i}$ and so
on.
Thus, one recognizes that the first two terms in above equation
describe the metric background of the model where the
metric in complex coordinates has the following block-diagonal
structure:
\be
G_{\mu\nu}=\pmatrix{0&K_{u_i\bar u_j}&0&0\cr K_{u_i\bar u_j}&0&0&0
\cr 0&0&0&-K_{v_p\bar v_q}\cr 0&0&-K_{v_p\bar v_q}&0\cr}.
\label{metric}
\ee
To obtain a space with Euclidean signature, one has to require that
$K_{u_i\bar u_j}$ is positive definite where
$K_{v_p\bar v_q}$ has to be negative definite.
Note that the metric (\ref{metric}) is not K\"ahler, whereas the
action
(\ref{action}) is nevertheless $N=2$ supersymmetric. If there are
only chiral resp. twisted chiral fields (i.e. $m=0$ or $n=0$)
one deals with a K\"ahler metric.

The $\epsilon_{ab}$-term in (\ref{bosonic}) provides the
antisymmetric
tensor field background:
\be
B_{\mu\nu}=\pmatrix{0&0&0&K_{u_i\bar v_p}\cr
0&0&K_{v_p\bar u_i}&0
\cr 0&-K_{v_p\bar u_i}&0&0\cr -K_{u_i\bar v_p}&0&0&0\cr}.
\label{bmn}
\ee
It follows that the field strength $H_{\mu\nu\lambda}$,
\be
H_{\mu\nu\lambda}
=\partial_\mu B_{\nu\lambda}
+\partial_\nu B_{\lambda\mu}
+\partial_\lambda B_{\mu\nu},\label{hfield}
\ee
can also be expressed entirely in terms
of the function $K$:
\be
\eqalign{
&H_{u_i\bar u_jv_p}
=
{\partial^3K\over\partial U_i\partial\bar U_j\partial V_p},\qquad
H_{u_i\bar u_j\bar v_p}
=-
{\partial^3K\over\partial U_i\partial\bar U_j\partial \bar V_p},\cr &
H_{v_p\bar v_qu_i}
=
{\partial^3K\over\partial V_p\partial\bar V_q\partial U_i},\qquad
H_{v_p\bar v_q\bar u_i}
=-
{\partial^3K\over\partial V_p\partial\bar V_q\partial \bar
U_i}.\cr}\label{
torsion}
\ee

So far we have just discussed the geometrical structure
of the $N=2$ backgrounds. In order that these backgrounds provide
consistent string solutions, they have to satisfy the string equation
of motion, i.e. the vanishing of the $\beta$-function equations.
In string theory, there is another background field, the dilaton
$\Phi(u_i,\bar u_i,v_p,\bar v_p)$,
which accompanies in general $G_{\mu\nu}$ and
$B_{\mu\nu}$. This amounts to add to the $\sigma$-model action
eq. (\ref{bosonic}) a term of the form ${1\over
2}R^{(2)}\Phi(u_i,v_p)$,
where $R^{(2)}$ is the scalar curvature of the two-dimensional
world-sheet.
Then the requirement of one-loop conformal
invariance of the two-dimensional $\sigma$-model
leads to the following equations of motion for the background fields,
\cite{beta}
\be
\eqalign{0&=\beta_{\mu\nu}^G=R_{\mu\nu}-{1\over
4}H_\mu^{\lambda\sigma}
H_{\nu\lambda\sigma}+2\nabla_\mu\nabla_\nu\Phi +O(\alpha')\cr
0&=\beta_{\mu\nu}^B=\nabla_\lambda
H_{\mu\nu}^\lambda-2(\nabla_\lambda
\Phi )H_{\mu\nu}^\lambda+O(\alpha').\cr}\label{betaf}
\ee
Moreover, the central charge deficit $\delta c$ provided by the
considered
background  is determined by the $\beta$-function of the dilaton
field as
\be
\delta c\equiv c-{3D\over 2}={3\over 2}\alpha'
\lbrack 4(\nabla\Phi)^2-4\nabla^2\Phi-R+{1\over 12}
H^2\rbrack+O(\alpha'^{2}).\label{dilaton}
\ee

We must emphasize here that in the presence of N=4 superconformal
symmetry the
solution to the lowest order in $\alpha$' is exact to all orders
in a specific scheme, and $\delta c$ remains zero to all orders.

\subsection{$U(1)$ duality transformations}
\setcounter{equation}{0}

Let us assume that a particular background, which satisfies
eqs. (\ref{betaf}) and (\ref{dilaton}) at lowest order in $\alpha'$,
really corresponds
to an exact conformal field theory. Then there exist
in many cases  different, so called dual, backgrounds which however
are
truly equivalent as a conformal field theory. In particular,
if the original background is independent from $d$ (real )
coordinates
$\theta_a$ ($a=1,\dots ,d$), i.e. if the original background
possesses
an Abelian
$U(1)^d$ isometry\footnote{Duality Symmetries for non-Abelian
isometries were investigated in ref. \cite{QO}.},
there exist dual but equivalent backgrounds
which are obtained from each other by discrete $O(d,d,{\bf Z})$
transformations \cite{GR}. (Moreover the field equations
(\ref{betaf}) and
(\ref{dilaton})
are invariant under continuous $O(d,d,{\bf R})$ transformations
\cite{odd}.
Using the continuous transformations
one can generate classes of backgrounds which
correspond to marginally deformed conformal field theories
\cite{K2,MAR}.)

Let us consider the most simple case of a single $U(1)$ isometry
assuming that the potential $K$ has one Killing symmetry and
is of the form
\be
K=K(Z+\bar Z,Y_i,\bar Y_i,V_p,\bar V_p)\label{killing}
\ee
where $Z$ and $Y_i$ are chiral fields, whereas $V_p$ are twisted
chiral
fields. (Of course the discussion holds in the same way if
$Z$ is a twisted chiral field.)

In \cite{LR,GHR} a duality transformation was described in which
twisted superfields are interchanged with untwisted ones.
Concretely, assume the existence of one
$U(1)$ Killing symmetry, $R=Z+\bar Z$,
and consider the `dual' potential
\be
\tilde K(R,Y_i,\bar Y_i,V_p,\bar V_p,\Psi+\bar\Psi)=
K(Z+\bar Z,Y_i,\bar Y_i,V_p,\bar
V_p)-R(\Psi+\bar\Psi),\label{dualpot}
\ee
where $Z$ is a chiral field and $\Psi$ a twisted chiral field.
Varying the action with respect to $\Psi$ gives back the original
theory. On the other hand one can equally well consider
the constraint coming from the variation with respect to $Z$,
\cite{LR}
\be
{\delta S\over\delta Z}=0\quad\rightarrow\quad
{\partial K\over\partial r}-(\psi+\bar\psi)=0,\label{constraint}
\ee
and
the dual theory is obtained as a Legendre
transform of $K$. Now the independent variable are $\psi$, $y_i$
and $v_p$.
It follows that the dual metric has the following form:
\be
\tilde G_{\mu\nu}=\pmatrix{
0&-\tilde K_{\psi\bar\psi}&0&0&0&-\tilde K_{\psi\bar v_q}\cr
-\tilde K_{\psi\bar\psi}&0&0&0&-\tilde K_{v_p\bar\psi}&0\cr 0&
0&0&\tilde K_{y_i\bar y_j}&0&0\cr 0&0& \tilde
K_{y_i\bar y_j}&0&0&0
\cr 0&-\tilde K_{v_p\bar\psi}&
0&0&0&-\tilde K_{v_p\bar v_q}\cr -\tilde K_{\psi\bar v_q}&0&
0&0&-\tilde K_{v_p\bar v_q}&0\cr}.
\label{dualmetric}
\ee
Similarly, the dual antisymmetric tensor field is obtained as
\be
B_{\mu\nu}=\pmatrix{0&0&0&\tilde K_{\psi\bar y_i}&0&0\cr
0&0&\tilde K_{y_i\bar\psi}&0&0&0\cr
0&-\tilde K_{y_i\bar\psi}&
0&0&0&\tilde K_{y_i\bar v_p}\cr -\tilde K_{\psi\bar y_i}&0&
0&0&\tilde K_{v_p\bar y_i}&0
\cr 0&0&0&-\tilde K_{v_p\bar y_i}&0&0\cr 0&0&
-\tilde K_{y_i\bar v_p}&0&0&0\cr}.
\label{dualbmn}
\ee

In \cite{RV} it was shown that this N=2 duality transformation via
Legendre
transform is the same as the usual abelian duality transformation
\cite{BU}.
This connection is important because one can compute also the
``quantum
corrections" to this classical duality transformation, which amount
to
a non-trivial dilaton.
We will discuss explicitly this connection, since the discussion
in \cite{RV} was not sufficiently general.

Introducing real coordinates as $z={r\over 2}+i\theta$
the bosonic action (\ref{bosonic}) becomes
\be
\eqalign{
S=&-{1\over 2\pi}\int{\rm d}^2x\Biggl\lbrack {K_{rr}\over
4}\partial^a r
\partial_a r+K_{rr}\partial^a\theta\partial_a\theta
+K_{y_i\bar y_j}\partial^ay_i\partial_a\bar y_j
\cr &+
K_{ry_i}({\partial^ar\over 2}-i\partial^a\theta)\partial_ay_i
+K_{r\bar y_i}({\partial^ar\over 2}+i\partial^a\theta)\partial_a\bar
y_i
-K_{v_p\bar
v_q}\partial^a v_p\partial_a\bar v_{ q}
\cr& +\epsilon_{ab}\biggl(
K_{r\bar v_p}({\partial_ar\over 2}+i\partial_a\theta
)\partial_b\bar v_p+K_{rv_p}({\partial_ar\over 2}-i\partial_a\theta)
\partial_bv_p\cr &+
K_{y_i\bar v_p}\partial_a y_i\partial_b\bar v_{ p}
+K_{v_p\bar y_i}\partial_a v_p\partial_b\bar y_{i}\biggr)
\Biggr\rbrack,\cr}
\label{bosonickill}
\ee
It is evident that the background does not depend
on the coordinate $\theta$. Thus one can perform a duality
transformation with respect to this $U(1)$ Killing symmetry,
and the dual metric and $B_{\mu\nu}$ are given as, \cite{BU}
\be\eqalign{&\tilde G_{\theta\theta}={1\over 2K_{rr}},\quad
\tilde G_{\theta v_p}={iK_{rv_p}\over 2K_{rr}},\quad
\tilde G_{rr}={K_{rr}\over 2},\cr
&\tilde G_{ry_i}={K_{ry_i}\over 2},\quad \tilde G_{y_iy_j}=
{K_{ry_i}K_{ry_j}\over 2K_{rr}},\quad \tilde G_{y_i\bar y_j}=
K_{y_i\bar y_j}-{K_{ry_i}K_{r\bar y_j}\over 2 K_{rr}},\cr
&\tilde G_{v_pv_q}=-{K_{rv_p}K_{rv_q}\over 2K_{rr}},\quad
\tilde G_{v_p\bar v_q}=
-K_{v_p\bar v_q}+{K_{rv_p}K_{r\bar v_q}\over 2 K_{rr}},\cr
&\tilde B_{\theta y_i}={K_{ry_i}\over 2K_{rr}},\quad\tilde B_{rv_p}=
-{K_{rv_p}\over 2},\cr &
\tilde B_{y_iv_p}={K_{ry_i}K_{rv_p}\over 2K_{rr}},\quad
\tilde B_{y_i\bar v_p}=-K_{y_i\bar v_p}+
{K_{ry_i}K_{r\bar v_p}\over 2K_{rr}}.\cr}\label{dualbackgr}
\ee
(The background fields $\tilde G_{\bar y_i\bar y_j}$ etc.
follow from complex conjugation.)

\noindent The dual dilaton field has the form
\be
2\tilde \Phi=2\Phi-\log 2K_{rr}.\label{dualphi}
\ee

We can now show that the duality above and the (supersymmetry
preserving)
Legendre transform amount to the same thing.
With the help of the constraint eq. (\ref{constraint}) we obtain
the following explicit expressions for the dual background:
\be
\eqalign{&\tilde K_{\psi\bar\psi}=-{1\over K_{rr}},\quad
\tilde K_{\psi\bar y_i}={K_{r\bar y_i}\over K_{rr}},\quad
\tilde K_{\psi\bar v_p}={K_{r\bar v_p}\over K_{rr}},\cr
&\tilde K_{y_i\bar y_j}=K_{y_i\bar y_j}-{K_{ry_i}K_{ry_j}\over
K_{rr}},\quad
\tilde K_{y_i\bar v_p}=K_{y_i\bar v_p}-{K_{ry_i}K_{r\bar v_p}
\over K_{rr}},\cr
&\tilde K_{v_p\bar v_q}=K_{v_p\bar v_q}-{K_{rv_p}K_{r\bar v_q}\over
K_{rr}}.\cr}\label{dualb}
\ee
Performing a coordinate transformation of the form
$2\psi=K_r+i\theta$ one can finally
show that the dual background eq. (\ref{dualbackgr})
agrees with the expressions in eqs. (\ref{dualmetric},\ref{dualbmn}).

Consider briefly the special case of a four-dimensional background
with one chiral field $Z$ and one twisted chiral field $V$ ($m=n=1$)
and assume that there is a $U(1)$ isometry with respect to
$Z$, i.e. $K=K(Z+\bar Z,V,\bar V)$. Then the dual potential
\be
\tilde K(\Psi+\bar\Psi,V,\bar V)=K(Z+\bar Z,V,\bar V)-(Z+\bar Z)(\Psi
+\bar\Psi)\label{dualpoten}
\ee
contains only twisted fields and is therefore
a true K\"ahler potential leading to a K\"ahler metric
which follows from eq. (\ref{dualmetric}). The Ricci tensor
is then determined to be (in complex notation, $i=\psi,v$)
\be
\tilde R_{i\bar j}=-\partial_i\partial_{\bar j}\log\det\tilde G_{i
\bar j}=
-\partial_i\partial_{\bar j}\log(-{K_{v\bar v}\over
K_{rr}}).\label{ricdual}
\ee
Thus the dual, four-dimensional K\"ahler space is
Ricci-flat if $K_{rr}\propto K_{v\bar v}$.

In summary, the ${\bf Z}_2$ duality transformation
can be expressed by a Legendre transformation in the
superspace action which replaces a chiral superfield by
a twisted chiral superfield. We will use this observation
in the following chapters to generate $N=2$ supersymmetric
backgrounds with torsion from torsion-free K\"ahler spaces.
Moreover we will also demonstrate that backgrounds
with torsion have as dual counterpart very interesting
K\"ahler spaces with isometries which are Ricci-flat
but non-compact.

\section{K\"ahler Spaces without Torsion and their Dual}
\setcounter{subsection}{0}
\subsection{K\"ahlerian backgrounds}
\setcounter{equation}{0}

It is already well known that if the torsion vanishes and there is no
dilaton field the condition that a $\s$-model has N=2 supersymmetry
is that the target space is K\"ahler, \cite{kahler}.
If a non trivial dilaton is present but no superpotential then there
are no
additional restrictions. It is interesting though that if a
superpotential is present then there are non-trivial constraints to
be satisfied.
Such constraints have the interpretation that there there exist
appropriate
screening charges (always required when a non-trivial dilaton is
present) which preserve the N=2 superconformal invariance.
We will return to this in the Appendix since it is of direct
relevance to
generalizations of LG models with non-trivial dilaton.

For the time being we start with a K\"ahler manifold specified
(locally)
by its K\"ahler potential $K(u_{i},\ub_{i})$ and a dilaton field
$\P$.
The metric is given in terms of the K\"ahler potential by the
standard formula
\be
G_{ij}=G_{{\bar i}{\bar j}}=0\;\;\;,\;\;\; G_{i{\bar
j}}=K_{u_{i}\ub_{j}}
\label{kmetric}
\ee
It is obvious that the metric is invariant under the so called
K\"ahler transformations of the potential
\be
K(u_{i},\ub_{i})\rightarrow K(u_{i},\ub_{i})+\Lambda(u_{i})+{\bar
\Lambda}(\ub_{i})\label{ktran}
\ee
Then the Ricci-tensor takes its well-known form
\be
R_{u_i\bar u_j}=-\partial_{u_i}\partial_{\bar u_j}
U\;\;,\;\; R_{u_iu_j}=
R_{\bar u_i\bar u_j}=0\label{ric}
\ee
 with $U=\log\det K_{u_i\bar u_j}={1\over 2}\log\det G$.

The only condition for conformal invariance is $\beta_{\mu\nu}^G=0$
which here implies
\be
\Phi={1\over 2}U+f(u_i)+\bar f(\bar u_i),\label{kahl}
\ee
and
\be
\nabla_{u_i}\partial_{u_j}\Phi=\nabla_{\bar u_i}\partial_{\bar u_j}
\Phi=0.\label{phikah}
\ee
where $f$ is an arbitrary holomorphic function.
In addition the central charge to one-loop is
$c=3n+\delta c$ where $n$ is the complex dimension and
\be
\delta c={3\over 2}e^{2\Phi}\square e^{-2\Phi}=3 K^{u_i\bar u_j}(4\partial_{u_i}\Phi
\partial_{\bar u_j}\Phi-2\partial_{u_i}\partial_{\bar
u_j}\Phi ).\label{dckah}
\ee
is the one-loop correction.
We have also included the contribution of the world-sheet fermions.

If one demands for enlarged $N=4$ world-sheet
supersymmetry, this implies that the K\"ahlerian space has to be
hyper-K\"ahler.
If the theory is also positive, then $\delta c=0$ from CFT arguments.
In such a case the Riemann tensor is self-dual and therefore the
space is Ricci-flat.
Ricci flatness and $\delta c=0$ implies constant dilaton.
However, we note that the hyper-K\"ahler condition is not
the only way to obtain $\delta c=0$; in fact we will provide
$N=4$ examples which are non-Ricci-flat and have non-constant
dilaton field. These examples will be presented later.

Let us see how far we can go in solving the conditions for conformal
invariance (\ref{kahl},\ref{phikah}).
The holomorphic double derivative equations on the dilaton can be
written
(after raising one index) as
\be
\partial_{u_i}\left(K^{u_j{\bar u_k}}\d_{\bar
u_k}\P\right)=0\;\;\;,\;\;\; 
K^{\bar u_i u_j}K_{u_j\bar u_k}={\delta^{i}}_{k}\label{df}
\ee
along with its complex conjugate.
The general solution of (\ref{df}) is
\be
\P(u,\ub)=K_{\ub^{i}}{\bar\xi}^{\bar
u_i}(\ub)+{\bar\rho}(\ub)\label{fa}
\ee
\be
\P(u,\ub)=K_{u^{i}}\xi^{u_i}(u)+\rho(u)\label{fb}
\ee
where $\xi^{i}$ ($\bar \xi^{\bar i}$) transforms as a vector under
analytic
(anti-analytic) transformations in order that $\P$ transform as a
scalar.
The two equations  (\ref{fa},\ref{fb}) for $\Phi$ imply a
compatibility
condition for $K$.

Here we have to distinguish two sub-cases.
First, if $\xi^{u_i}={\bar \xi}^{\bar u_i}=0$
then $\rho={\bar\rho}=$constant and
we are
reduced to the usual CY case where the K\"ahler manifold has to be
Ricci flat.
We will return with many solutions to this case later on.
If however $\xi^{u_i}$,${\bar \xi}^{\bar u_i}$ are non-zero
then we will show that the dilaton is always non-trivial and that
there is
a generic Killing symmetry in the K\"ahler metric as well as the
dilaton.
To do this we observe that by an analytic (anti-analytic)
transformation of
the coordinates we can always rotate
$\xi^{u_i}$ (${\bar \xi}^{\bar u_i}$)
into a single direction and make its value there equal to one.
This has also the effect that it redefines the (otherwise) arbitrary
holomorhic function $f$ in (\ref{kahl}).
Thus, without loss of generality we can write the equivalent
conditions
\be
\P=K_{\bar z}+{\bar
\rho}(\bar z, \bar y_{i})=K_{z}+\rho(z,y_{i})\label{fc}
\ee
along with (\ref{kahl}), where $z$,$\bar z$ are the coordinates along
the preferred
directions and $y^{i}$,$\bar y^{i}$ denote the rest of the
coordinates.

It is already obvious from (\ref{fc}) that
\be
\d_{z}\P =\d_{\bar z}\P \; \rightarrow
\P=\P(z+\bar z,y_{i},\bar y_{i})\label{fd}
\ee
One can show that the compatibility of the equations (\ref{kahl}) and
(\ref{fd})
along with our
freedom to perform K\"ahler transformations implies that
\be
K=K(z+\bar z,y_{i},\bar y_{i})\;\;,\;\;\P=\d_{z}K=\d_{\bar z}
K\label{fe}
\ee
and
\be
\P={1\over 2}U+\eta(z+\bar z)\label{ff}
\ee
where $\eta$ is a real number.
We can take (\ref{ff}) as the equation specifying the dilaton in
terms of the
metric and then (\ref{ff}) becomes a non-linear differential equation
for the
K\"ahler potential
\be\det [K_{u_{i}\ub_{j}}]=\exp
[-2\eta(z+\bar z)+K_{z}+K_{\bar z}]
\label{fg}
\ee
generalizing the CY condition.
It should be kept in mind that this equation holds only in the
special
coordinate system defined above.
In the same coordinate system, specified above, we can also compute
the
central charge deficit:
\be
\delta c=12 \eta.\label{centralcq}
\ee

Let us consider a special class
of solutions which can be regarded as the
generalization of the Euclidean two-dimensional black-hole backgrounds
found in \cite{bh}. Specifically, assume that the model
has a $U(n)$ isometry, i.e.
\be
K=K(x),\quad \Phi=\Phi(x) \qquad
x=\sum_{i=1}^N|u_i|^2.\label{unansatz}
\ee
The general form of the metric is then
\be
K_{u_i\bar u_j}=K'\delta_{ij}
+K''\bar u_i u_j,\qquad K'={\partial K\over\partial
x}.\label{unisom}
\ee
and
\be
K^{\bar u_i u_j}={1\over K'}\left(\delta_{ij}-{K''\over K'+xK''}\bar u_iu_j\right)
\ee

For $n>1$, the linear term in the dilaton, eq. (\ref{ff}),
is not allowed by the $U(n)$ isometry and the dilaton field becomes
\be
\P={1\over 2}U={1\over 2}\log
\lbrack(K')^{n-1}(K'+xK'')\rbrack.\label{dilisom}
\label{neweq1}\ee
Let us define the following function:
\be
Y(x)=xK'(x).\label{yfct}
\ee
Now we have to insert the ansatz eqs. (\ref{unisom}, \ref{dilisom})
into the field equation (\ref{phikah}), to obtain
\be
\Phi''+{2K''+xK'''\over K'+xK''}\Phi'=0
\ee
which can be integrated twice to obtain
\be
\Phi=-{n\over 2k}xK'+C
\label{neweq}\ee
where $k,C$ are real constants. 
Together (\ref{neweq1}) and (\ref{neweq}) imply
\be
Y^{n-1}~e^{nY\over k}~ Y'=x^{n-1}e^{2C}
\ee
The general solution of this
equation takes the following form:
\be
e^{nY\over k}\sum_{m=0}^{n-1}{(-1)^m\over m!}\left({nY\over k}\right)^m=A+Bx^n.\label{solution}
\ee
Here $A$, $B$, $k$ are arbitrary real constants.

{}From  (\ref{solution}) we immediately obtain
\be
Y'=kB(n-1)!(-1)^{n-1}e^{-{nY\over k}}\left({nY\over k}\right)^{1-n}x^{n-1}.\label{yprime}
\ee
Then the dilaton, eq. (\ref{dilisom}), can be also expressed
entirely of $Y$ as
\be
\P={1\over 2}U=-{n\over 2k}Y+{\rm const.}\label{dily}
\ee
The Ricci tensor becomes
\be
R_{u_i\bar u_j}=-
\partial_{u_i}\partial_{\bar u_j}U=
Y'\delta{ij}+Y''\bar u_i u_j.\label{ricciy}
\ee
The scalar curvature can be computed to be
\be
R=2(n-xY')=2(n-f_n(Y))
.\label{scalarcy}
\ee
with
\be
f_n(Y)=k\left({nY\over k}\right)^{1-n}\biggl\lbrack
\sum_{m=0}^{n-1}{(-1)^m\over m!}\left({nY\over k}\right)^m-Ae^{-{nY\over k}}\biggr\rbrack (-1)^{n-1}(n-1)!
\ee
Finally, the one-loop corrected central charge of the supersymmetric theory is 
\be
c=3n+6{n^2\over k}.\label{centralun}
\ee

Making a coordinate change 
\be
u_i=\sqrt{x}~e^{i{\theta\over 2}}~z^i \;\;,\;\; z_i={y^i\over \sqrt{1+\sum_{i=1}^{N-1}y^{i}\bar
y^{i}}}\;\;,\;\; z^N={1\over \sqrt{1+\sum_{i=1}^{N-1}y^{i}\bar
y^{i}}}
\ee
and introducing the real coordinate $Y$
defined in
(\ref{yfct}), the metric becomes.
\be
\dd s^2={(\dd Y)^2\over 4f_n(Y)}+{f_n(Y)\over 4}\left[\dd
\theta-i\left(\Gamma_{i}\dd  y^{i}-\Gamma_{\bar i}\dd \bar
y^{i}\right)\right]^2
+Yg_{i\bar j}\dd y^{i}\dd \bar y^{j}
\ee
with
\be
\Gamma_{i}=\partial_{i}\log(1+\sum_{i=1}^{N-1}y^{i}\bar
y^{i})\;\;,\;\;g_{i\bar
j}=
\partial_{i}\partial _{\bar j}\log(1+\sum_{i=1}^{N-1}y^{i}\bar y^{i})
\ee
$g_{i\bar j}$ is the invariant Fubini-Study metric of the coset ${SU(N)\over
SU(N-1)\times
U(1)}\sim CP^n$. The angular variable $\theta$ has period $4\pi$ as 
evident from its definition.

The explicit form of the scalar curvature, eq. (\ref{scalarcy}),
allows
us to discuss the asymptotic behavior and the singularity
structure of our class of solutions. First we recognize
that the $2n$-dimensional K\"ahler space
has zero scalar curvature for $Y\rightarrow \infty$.
Moreover, there is another (conical) singularity at $Y=0$ for $n>1$
if $A\neq 1$. On the other hand, if $A=1$, the manifold is regular.

Let us study first the simplest, already well-known
case, namely the two-dimensional
backgrounds.
First assume that the dilaton field is of the form $\Phi={1\over
2}U$.
Then we obtain from eq. (\ref{solution}) that
\be
Y=-2k~\Phi={k\log(A+Bx)}.\label{solna}
\ee
The corresponding K\"ahler potential is obtained by integrating
$Y/x$,
and it is given by the dilogarithm like
\be
K(x)=k\int_1^{Bx}{{\rm
d}\alpha\over\alpha}\log(A+\alpha).\label{dilog}
\ee
The backgrounds are singular for $x=-A/B$ ($A\neq 0$).
In fact, inequivalent background metrics are
just characterized by the sign of $A$ and $B$ and we distinguish the
following
four cases ($K_{u\bar u}=Y'$):
\be
({\rm i})\qquad A=0\;{\rm and}\; B=1\qquad K_{u\bar u}={k\over ~u\bar
u}.\label{casea}
\ee
After a change of coordinates, $u=e^z$, the metric becomes flat,
${\rm d}s^2=k~{{\rm d}z{\rm d}\bar z}$, together with a linear
dilaton of the form $\P\sim (z+\bar z)$.

\be
\eqalign{&
({\rm ii})\qquad B>0,\qquad A<0\qquad K_{u\bar u}={k\over (u\bar
u-1)},\cr &
({\rm iii})\qquad B<0,\qquad A>0\qquad K_{u\bar u}={k\over (1- u\bar
u)}.\cr}
\label{caseb}
\ee
The background (ii)
corresponds to the exact conformal field theory based on the
gauged WZW model with coset $SL(2,{\bf R})/U(1)_{\rm vector}$.
This is the so called Euclidean black hole with the form of a
trumpet.
Here the central charge deficit is positive:
$\delta c=6/k$.
For case (iii), the corresponding conformal field theory
belongs to the gauged WZW model with coset $SU(2)/ U(1)$,
where the background space is now compact with positivity domain
$0\leq x<1$. Now the central charge deficit is negative:
$\delta c=-6/k$.
Performing an analytic continuation from the Euclidean
to Minkowski space-time, one deals with the two-dimensional
black-hole for (ii), but with a cosmological scenario for (iii).

Finally
\be
({\rm iv})\qquad A>0,\qquad B>0\qquad K_{u\bar u}={1\over C(u\bar
u+1)}.\label{casec}
\ee
Now, the non-compact background space with cigar shape
corresponds to
the exact conformal field theory that is described by a coset
$SL(2,{\bf R})/ U(1)_{\rm axial}$.
The central charge deficit is positive:
$\delta c=6/k$.
After analytic continuation
the background has  again a black-hole  interpretation.

As already mentioned, in two dimensions the dilaton field may
have a linear contribution and is in general given as (cf.eq.
(\ref{ff})):
\be
\Phi={1\over 2}U+q\log x.\label{dillin}
\ee
On the other hand eq. (\ref{fc}) requires that $\Phi\propto Y$
($K_u=Y$
with $x=u\bar u=e^{z+\bar z}$),
and with eq. (\ref{dillin}) we obtain the following differential
equation
\be
-Y=\log Y'+2q\log x.\label{diffy}
\ee
For $q\neq {1\over 2}$ the solution to (\ref{diffy}) has the form
$Y=\log(x^{1-2q}+c)-\log(1-2q)$. This just corresponds
to the already found solutions (i) - (iv).
On the other hand, for $q={1\over 2}$ a new solution
is found:
\be
Y=\log\log x.\label{newsol}
\ee
The corresponding background metric then looks like
\be
({\rm v})\qquad K_{z\bar z}={k\over (z+ \bar z)},\qquad .\label{cased}
\ee
As we will show later, this solution is just the dual of the
flat background with constant dilaton field, \cite{K2}.
Note that cases (i) -- (v) exhaust all
possible solutions in two dimensions (except the trivial one, namely
flat space with constant dilaton).

Now let us switch to the four-dimensional backgrounds.
Obviously, four-dimensional backgrounds with $N=2$ supersymmetry
are obtained by building all possible tensor products \cite{C2}
of the two-dimensional cases listed just before.
This gives 21 possibilities where, however, several cases are
equivalent since they are the dual of each other.
In addition it is interesting to note that if one couples
the cosmological space (ii), based on the coset $SU(2)\over U(1)$
with level $k_1$, to any other two-dimensional solution
with level $k_2$, then the resulting central charge defect vanishes,
i.e. $\delta c=0$ provided if $k_1=k_2$.
(In the corresponding coset conformal field theories,
the level must be related like $k_2-k_1=4$ for $SU(2)/U(1)_{k_1}
\otimes SU(2,{\bf R})/U(1)_{k_2}$.)
In fact these types of product spaces lead
to an enhanced $N=4$ world-sheet supersymmetry. The corresponding
$N=4$ superconformal algebras were explicitly constructed in ref.
\cite{C2}.

Let us now consider four-dimensional backgrounds which are not direct
products of two-dimensional spaces. Specifically we consider
solutions of the form eq. (\ref{solution}) with
$n=2$. One has to emphasize that so far it is not known to us
which exact superconformal field theory might correspond to
this type of backgrounds.
Eq. (\ref{solution}) reads
\be
e^{2Y\over k}\left(1-{2\over k}Y\right)=A+Bx^2.\label{solutntwo}
\ee
The solution of this equation can be expressed in term of
the (inverse) incomplete $\gamma$-function $W(x)$:
\be
Y(x^2)={k\over 2}W\biggl(-{A+Bx^2\over e}\biggr)+1.\label{incomplg}
\ee

In order to discuss the properties of the obtained four-dimensional
K\"ahler spaces for generic choices of the parameters $A$ and $B$,
it is convenient to switch to real coordinates, one of them being
$Y$.
The metric then reads:
\be
\eqalign{{\rm d}s^2&={({\rm d}Y)^2\over 4f_2(Y)}+{f_2(Y)\over 4}({\rm d}\theta
+\cos\psi{\rm d}\phi)^2
+{Y\over 4}\biggl(({\rm d}\psi)^2+\sin^2\psi({\rm d}\phi)^2\biggr),\cr
&f_2(Y)=xY'={k^2\over 2Y}\left(Ae^{-{2Y\over k}}+{2\over k}Y-1\right).\cr}\label{betterme}
\ee

The scalar curvature is
\be
R=4\biggl(1+{k^2\over 4}{1-{2\over k}Y-Ae^{-{2Y\over k}}\over Y}\biggr).\label{scalarctwo}
\ee

In these coordinates the $SU(2)\times U(1)$ killing symmetry
of the metric is also manifest.
The $SU(2)$ symmetry acts on the coordinates
$\psi,\phi$ and translates also $\theta$,
\be
\delta^{3}\phi=\varepsilon\;\;,\;\;\delta^{3}\theta=
\delta^{3}\psi=0\label{j3}
\ee
\be
\delta^{\pm}\phi={\varepsilon^{\pm}\over 2}\cot \psi e^{\mp
i\phi}\;\;,\;\;
\delta^{\pm}\psi=\pm i{\varepsilon^{\pm}\over 2}e^{\mp
i\phi}\;\;,\;\;
\delta^{\pm}\theta=-{\varepsilon^{\pm}\over 2}{e^{\mp i\phi}\over
\sin \psi}
\ee
The $U(1)$ acts as translations on $\theta$ only.
The metric (\ref{betterme}) in the ($\theta,\psi,\phi$)
subspaces is a deformation of the fibration of $S^{3}$ over
$S^2$ (whose line element is manifest in (\ref{betterme}))
with fiber line element ${\rm d}\theta
+\cos\psi{\rm d}\phi$.

Since this string solution has two abelian killing symmetries we can
use
O(2,2,R) transformations to generate a family of solutions.
However, a priori O(2,2,R) transformations may break the N=2
superconformal
invariance.
Identifying in general such an N=2 subgroup is a difficult problem
and we will not attempt it here.
We will indicate however a subgroup that always preserves  N=2.
This is generated by the abelian duality transformations described in
section 2.2 and constant antisymmetric tensor shifts (in general
gauge transformations of $B_{\mu\nu}$ that leave $H_{\mu\nu\rho}$
invariant, i.e. $B_{\mu\nu}\to B_{\mu\nu}+\partial_{\mu}\Lambda_{\nu}
-\partial _{\nu}\Lambda_{\mu}$).
Abelian duality transformations preserve N=2 invariance.
Constant $B_{\mu\nu}$ shifts (or in general gauge transformations of
B)
also preserve N=2 supersymmetry since the extra terms in the action
are total derivatives (although they change the spectrum in general).

$B_{\mu\nu}$ gauge transformations are generated (in the case where
$H\not= 0$) by the quasi-K\"ahler
transformations (\ref{inva}).
These can be intertwined with the U(1) duality transformations to
generate
non-trivial orbits of solutions since B-gauge transformations and
U(1)
duality do not commute.

Not all such B-shifts can be done in an N=2 invariant language, (for
example
a constant B-shift in the K\"ahler case).
It is an interesting problem to describe all duality transformations
which contain $B_{\mu\nu}$ gauge transformations
in a manifestly N=2 invariant fashion.

Thus, the generic duality group (subgroup of O(d,d)) which preserves
N=2 is generated by the d $Z_{2}$ duality transformations $D_{i}$
and constant B-shifts (in Killing coordinates).
The $D_{i}$ act as the O(d,d) matrices
\be
D_{i}=\left(\matrix{e_{i}&1-e_{i}\cr 1-e_{i}&e_{i}\cr}\right)
\ee
where $e_{i}$ is a $d\times d$ diagonal matrix with all elements zero
along the diagonal except the i-th one which is equal to one.
The transformations $D_{i}$ satisfy
\be
D_{i}^{2}=1\;\;\,\;\;D_{i}D_{j}=D_{j}D_{i}
\ee
B-shifts constitute an additive $d(d-1)/2$-dimensional abelian group
which acts as
\be
D_{B}=\left(\matrix{1&{\bf B}\cr 0&1\cr}\right)
\ee
where ${\bf B}$ is a $d\times d$ antisymmetric matrix.

In the case of our solution, we can obtain a 1-parameter family of
deformations.
Here we will take the O(2,2) matrix to have the form
\be
\left(\matrix{1&0&0&0\cr
 0&1&0&0\cr
0&\lambda&1&0\cr
-\lambda&0&0&1\cr}\right)=\left(\matrix{0&0&1&0\cr
0&0&0&1\cr 1&0&0&0\cr 0&1&0&0}\right)\left(\matrix{
1&0&0&\lambda\cr 0&1&-\lambda&0\cr
0&0&1&0\cr 0&0&0&1\cr}\right)
\left(\matrix{0&0&1&0\cr
0&0&0&1\cr 1&0&0&0\cr 0&1&0&0}\right)
\label{o22}
\ee
where $\left(\matrix{0&1\cr 1&0\cr}\right)=D_{1}D_{2}$.
This will miss some discrete transformations which will be dealt
with in the next subsection.
The metric is given by
\be
4{\rm d}s^2={({\rm d}Y)^2\over f_2(Y)}+Y({\rm d}\psi)^2+{f_2(Y)({\rm
d}\theta
+\cos\psi{\rm d}\phi)^2+Y\sin^2\psi ({\rm d}\phi)^2\over
1+\lambda^{2}Yf_2(Y)\sin^2\psi}
\label{defor}
\ee
where $\lambda$ is a real number and $f_2(Y)$ was defined in
(\ref{betterme}). At
$\lambda=0$ we obtain our original solution, (\ref{betterme}).
Now $B_{\mu\nu}$ is non-zero:
\be
B_{\theta\phi}=-{\lambda Yf_2(Y)\sin^2\psi\over
1+\lambda^{2}Yf_2(Y)\sin^2\psi}
\label{B}
\ee
Finally the dilaton is given by
\be
\Phi=-{1\over
2}\log[e^{2Y\over k}(1+\lambda^{2}Yf_2(Y)\sin^2\psi)]\label{dedil}
\ee
The family of manifolds described by (\ref{defor}) is asymptotically
flat
as $Y\rightarrow\infty$ (except when $\lambda=0$) and regular when
$f_2(Y)$
and $Yf_2(Y)$ are positive.

We will analyse the structure of the Euclidean manifold as a function
of A and B.
We need some asymptotics of the function $f_2(Y)$:
\be
f_2(\infty)=2 \;\;,\;\;f_2(-\infty)=-{\rm Sign}[A]\times \infty\label{f1}
\ee
\be
f_2(0^{+})=-f_2(0^{-})={\rm Sign}[A-1]\times \infty\label{f2}
\ee
Since $x^2$ must be positive, we are dealing with the following
cases:

1) $A>1,B<0$. There are two manifolds, the first with $Y\geq 0,
x^2>(1-A)/B$
with signature (4,0)
and a curvature singularity at $Y=0$ and the second with$Y\leq 0,
(1-A)/B\leq x^2 \leq -A/B$ with signature (0,4)
and curvature singularities at $Y=0,-\infty$.

2) $A=1, B<0$. There is a regular Euclidean (4,0) manifold for $Y>0,
x^2 >0$
and a
singular (at $Y=-\infty$) Eclidean (0,4) manifold for $Y<0, 0\leq x^2
\leq
-1/B$.

3) $0<A<1$. In this case $f_2(Y)$ has a positive and a negative zero
which
we will denote by $Y_{\pm}$: $f_2(Y_{\pm})=0$.
For $B<0$ there is again a regular (finite curvature) Euclidean
manifold for
$Y>Y_{+}$ with signature (4,0) and another with $Y<Y_{-}$ with
signature (0,4)
and a curvature singularity at $Y=-\infty$.
For $Y_{-}<Y<Y_{+}$ and $B>0$ there is another singular  manifold
with
signature (2,2).

4) $A\leq 0$. In this case $f_2$ has a single positive zero, $Y_{+}$.
For $B<0$ and $Y>Y_{+}$ we have a regular manifold with signature
(4,0).
For $B>0$ and $Y<Y_{+}$ there is a singular manifold with signature
(2,2).

\subsection{Dual spaces}
\setcounter{equation}{0}

Having found explicit K\"ahlerian backgrounds which solve the
string equations of motion we can now construct their dual
spaces replacing one chiral field $u$ by a twisted chiral field
$\psi$.
Then the dual space will in general contain non-trivial torsion and
the metric is no longer K\"ahlerian.
For simplicity we assume again that $K=K(x)$,
$x=\sum_{i=1}^N|u_i|^2$.
We will perform the duality transformation with respect to the
overall $U(1)$ isometry. Thus we define a change of coordinates
like follows $x=u_1\bar u_1(1+\sum_{i=1}^{N-1}y_i\bar y_i)=u_1\bar
u_1
(1+x')$, and the dual potential has the form
\be
\tilde K=K(x)-(\psi+\bar\psi)\log{x\over 1+x'}.\label{dualun}
\ee
{}From this eq. (\ref{constraint}) becomes
\be
Y=\psi+\bar\psi ,\label{constry}
\ee
and the dual metric is then determined by
\be
\tilde K_{\psi\bar\psi}=-{1\over x}{\partial x\over \partial \psi},
\quad \tilde K_{y_i\bar y_j}=(\psi+\bar\psi)\biggl\lbrack
{\delta_{ij}\over 1+x'}-{y_j\bar y_i\over
(1+x')^2}\biggr\rbrack.\label{dualmsun}
\ee
We recognize the `transverse' metric $\tilde K_{y_i\bar y_j}$
is exactly the K\"ahlerian metric of the complex sphere
$SU(n)/(SU(n-1)\times U(1))$ with `radius' $\psi+\bar\psi$.
The antisymmetric tensor simply becomes
\be
\tilde K_{\psi\bar y_i}={y_i\over 1+x'}.\label{dualbmnsu}
\ee

The simplest example of this type is provided by the dual
of flat Euclidean space and constant dilaton $\Phi$.
Here $K(x)=x$ and therefore
$\tilde K_{\psi\bar\psi}=-{1\over\psi+\bar\psi}$.
Thus the dual of flat space is singular at $\psi+\bar\psi=0$.
Performing a change of coordinates like $\psi={t^2\over 4}+i{x\over
2}$
the metric has the form
\be
{\rm d}\tilde s^2={\rm d}t^2+{1\over t^2}{\rm d}x^2+{t^2}
\biggl\lbrack
{\delta_{ij}\over 1+x'}-{y_j\bar y_i\over (1+x')^2}\biggr\rbrack
{\rm d}y_i{\rm d}\bar y_j.\label{dualflat}
\ee
The dual dilaton field has the form:
\be
\tilde\Phi=-{1\over 2}\log(\psi+\bar\psi)=-\log t+{\rm constant}.
\label{dilpsi}
\ee
Thus, in two dimensions we have just found our previous solution
(iv),
eq. (\ref{cased}).
It is interesting to note that the Einstein metric in 4-d
corresponding
to (\ref{dualflat}, \ref{dilpsi}) is flat.
As we will see later,this solution is in a class of N=4 solutions
which are axionic-dilatonic instantons.

Now consider the non-trivial K\"ahler spaces with metrics determined
by eq. (\ref{solution}).
Using eq. (\ref{dualmsun}), the dual metric becomes
($Y=\psi+\bar\psi$)
\be
\tilde K_{\psi\bar\psi}={1\over
f_n(Y)}.\label{dualsolution}
\ee
For $n=1$, the dual metric looks like
\be
{\rm d}\tilde s^2={k\over (z\bar z-A)}{\rm d}z{\rm d}\bar z,
\quad (z=e^{\psi/k}).\label{dualnone}
\ee
Comparing with eqs. (\ref{caseb},\ref{casec}), it is evident that the
duality
transformation maps the Euclidean black-holes, $k>0$,
with cigar and
trumpet forms, respectively, onto each other,\cite{Gi,DVV,K1},
whereas the
cosmological
solution, $k<0$, is self-dual.

For the four-dimensional case with $n=2$, the dual metric is given as
\be
{\rm d}\tilde s^2={
{\rm d}\psi{\rm d}\bar\psi\over f_2(Y)}
+{Y\over (1+y\bar y)^2}{\rm d}y{\rm d}
\bar y,
\label{dualntwo}
\ee
whereas the dual dilaton and antisymmetric tensor field look like
\be
\tilde\Phi=-{1\over 2}\log[e^{2Y\over k}f_2(Y)],\qquad \tilde B_{\psi
\bar y}={2y\over 1+y\bar y}.\label{dilbmn}
\ee
The dual scalar curvature becomes
\be
\tilde R ={4Y^2(f_2f_2''-f_2'^{2})+2f_2^{2}+8Yf_2\over Y^2 f_2}\label{dualsc}
\ee
Thus, for  the dual space, $\tilde R =0$  for $Y\rightarrow\infty$,
and there are curvature singularities at $Y=-\infty,0$ and, for
generic values
of $A$ at the zeros of $f_2(Y)$.

This solution also can be boosted with the N=2 preserving O(2,2,R)
transformations mentioned in the previous section.

It is illustrative to consider also the dual of a four-dimensional
space which is a direct product of two two-dimensional subspaces.
Specifically take the tensor product of two-dimensional flat
space with the cosmological solution (iii).
The starting K\"ahler potential has the form
\be
K^{total}=K(x)+{1\over 2}(u_2+\bar u_2)^2.\label{tensor}
\ee
Here $Y(x)=xK'=-\log(1-x)$.
We performed a change of coordinates like $u_2=-\log u_1+\log y$, and
the dual potential is obtained as
\be
\tilde K^{total}=K(x)+{1\over 2}(\log x-\log y\bar
y)^2-(\psi+\bar\psi)\log x.
\label{dualtensor}
\ee
Then, after the standard algebra the dual metric is found as
($z=e^{-\psi}
$):
\be
{\rm d}\tilde s^2={{\rm d}z{\rm d}\bar z+{\rm d}y{\rm d}\bar y
\over z\bar z+y\bar y}.\label{wormhole}
\ee
It is not difficult to show that this is exactly the metric
of the so called semi wormhole space \cite{Ca,C1,C2}. (The inverse
duality
was first discovered in \cite{RSS}.)
It corresponds to the exact
conformal field theory, based on the WZW model $SU(2)\times U(1)$.
So after a suitable projection in the Hilbert space,
the conformal field theory, based on $SU(2)/U(1)\otimes U(1)^2$
becomes equivalent to $SU(2)\times U(1)$ conformal field theory
\cite{C2}.
It is not difficult to show that the torsion which arises after
the duality transformation is just given by the structure
constants of $SU(2)$.

\section{Non-K\"ahlerian Spaces with Torsion}
\setcounter{subsection}{0}
\subsection{Four-dimensional non-K\"ahlerian spaces}
\setcounter{equation}{0}

To start with
non-vanishing antisymmetric tensor fields we restrict ourselves
to the simplest case, namely to
four-dimensional target-spaces, i.e. $m=n=1$.
Then, defining $U=\log K_{u\bar u}$ and $V=\log K_{v\bar v}$,
the Ricci-tensor becomes
\be
\eqalign{R_{uu}&=-\partial_u\partial_uV-{1\over 2}\partial_uV
\partial_uV+\partial_uV\partial_uU,\cr
R_{u\bar u}&=-\partial_u\partial_{\bar u}U+{1\over 2}\partial_uV
\partial_{\bar u}V,\cr
R_{uv}&=-{1\over
2}(\partial_u\partial_v(U+V)-\partial_uV\partial_vU),
\cr
R_{u\bar v}
&=-{1\over 2}(\partial_u\partial_{\bar v}(U+V)-\partial_uV
\partial_{\bar v}U).\cr}\label{riccitor}
\ee
(The remaining components of the Ricci-tensor follow from symmetry
arguments and complex conjugation.)
Analogously, the tensor $F_{\mu\nu}=H_{\mu\rho\sigma}H_{\nu\lambda
\kappa}G^{\rho\lambda}G^{\sigma\kappa}$ has the form
\be
\eqalign{F_{uu}&=-2\partial_uV\partial_uV,\cr
F_{u\bar u}&=2\partial_uV\partial_{\bar u}V-4e^{U-V}
\partial_vU\partial_{\bar v}U,\cr
F_{uv}&=-2\partial_uV\partial_vU,
\cr
F_{u\bar v}
&=-2\partial_uV\partial_{\bar v}U
.\cr}\label{hh}
\ee
Then the metric $\beta$-functions is computed to be
\be
\eqalign{\beta_{uu}^G&=-\partial_u^2(V-2\Phi)+\partial_uU\partial_u
(V-2\Phi),\cr
\beta_{u\bar u}^G&=-\partial_u\partial_{\bar u}(U-2\Phi)+e^{U-V}
\partial_vU\partial_{\bar v}(U-2\Phi),\cr
\beta
_{uv}^G&=-{1\over 2}\partial_u\partial_v(U+V-4\Phi
)-\partial_u\Phi\partial_vU-\partial_v\Phi\partial_u V+\partial_uV
\partial_vU,\cr
\beta
_{u\bar v}^G&=-{1\over 2}\partial_u\partial_{\bar v}(U+V-4\Phi
)-\partial_u\Phi\partial_{\bar v}U
-\partial_{\bar v}\Phi\partial_u V+\partial_uV
\partial_{\bar v}U.\cr}
\label{betamtor}
\ee
For $\beta_{\mu\nu}^B$ we get
\be\eqalign{
\beta_{u\bar u}^B&=\partial_u\Phi\partial_{\bar u}V-
\partial_uV\partial_{\bar u}\Phi,\cr
\beta
_{uv}^B&=\partial_u\partial_v(U-V
)-2(\partial_u\Phi\partial_vU-\partial_v\Phi\partial_u V)
,\cr
\beta
_{u\bar v}^B&=\partial_u\partial_{\bar v}(U-V
)-2(\partial_u\Phi\partial_{\bar v}U
-\partial_{\bar v}\Phi\partial_u V).\cr}
\label{betabtor}
\ee
Finally the central charge deficit becomes
\be
\eqalign{
\delta c={3\alpha' \over 2}\lbrace e^{-U}
(8\partial_{u}\Phi
\partial_{\bar u}\Phi-2\partial_{u}\partial_{\bar u}U
-4\partial_uV\partial_{\bar u}V)\cr
-e^{-V}
(8\partial_{v}\Phi
\partial_{\bar v}\Phi-2\partial_{v}\partial_{\bar v}V
-4\partial_vU\partial_{\bar v}U)\rbrace .\cr}
\label{dckahtor}
\ee

Let us now try to solve the
$\beta$-function equations (\ref{betamtor}) and (\ref{betabtor}).
First, the vanishing of $\beta_{uu}^G$, $\beta_{vv}^G$,
$\beta_{uv}^G$ and $\beta_{uv}^B$ implies that
\be
\eqalign{&\partial_uV=2\partial_u\Phi+\bar C_1(\bar u,\bar v)e^U,\cr
&\partial_vU=2\partial_v\Phi+\bar C_2(\bar u,\bar v)e^V.\cr}\label{a}
\ee
Analogous equations hold for the derivatives with respect to $\bar u$
and $\bar v$. Using $\beta_{u\bar v}^G=\beta_{u\bar v}^B=0$
it follows that $\partial_uC_2=\partial_{\bar v}\bar C_1=0$.
Thus eqs. (\ref{a}) can be rewritten as
\be
\eqalign{&\partial_uV=2\partial_u\Phi+\bar C_1(\bar u)e^U,\cr
&\partial_vU=2\partial_v\Phi+\bar C_2(\bar v)e^V.\cr}\label{b}
\ee

In general, in order to proceed we discuss separately the following
exclusive cases:
(i) $C_1=C_2=0$; (ii) $C_1=0$, $C_2\neq 0$; (iii) $C_1,C_2\neq 0$.
Case (i) implies
\be
U-2\Phi={\rm constant},\qquad V-2\Phi={\rm constant}.\label{i}
\ee
Thus $U-V={\rm constant}$.
Indeed, case (i) leads to $\delta c=0$
and corresponds to a $N=4$ supersymmetric background. We
will extensively discuss this case  at the end of this chapter.

For the case (ii) it is very useful to perform the following
change of coordinates:
\be
w=\int{{\rm d}v\over C_2(v)}.\label{changew}
\ee
Then the condition $\beta_{v\bar v}^B=0$ implies $\Phi=\Phi(u,\bar u,
w+\bar w)$. In addition, eqs. (\ref{b}) tell us that $V=V(u,\bar
u,w+\bar w)$
and $U=U(u,\bar u,w+\bar w)$. Thus this case necessarily leads to at
least one
$U(1)$ Killing symmetry.
In summary, for case (ii) the background is determined by the
following
differential equations:
\be
\eqalign{&\partial_u(V-2\Phi)=\partial_{\bar u}(V-2\Phi)=0,\cr
&\partial_w(U-2\Phi)=e^V,\cr
&\partial_w^2(V-2\Phi)=0,\cr
&\partial_u\partial_{\bar
u}(U-2\Phi)=\partial_we^U.\cr}\label{oneiso}
\ee
The last two equations follow from the vanishing of
$\beta_{u\bar u}^G$ and $\beta_{v\bar v}^G$.
These equations can be solved by
\be
V-2\Phi=c_1(w+\bar w)+c_2,\label{solvea}
\ee
where this equation defines the dilaton. The potential $K$ has to
satisfy
\be
K_{ww}=K_{u\bar u}e^{-K_w+c_1(w+\bar w)+c_2}.\label{solveb}
\ee
where the central charge deficit $\delta c$ is proportional to the
constant
$c_{1}$ appearing in (\ref{solveb}).

Switching now to case (iii), we change, in addition to
(\ref{changew}),
the coordinate $u$ as follows
\be
z=\int{{\rm d}u\over C_1(u)}.\label{changez}
\ee
Furthermore it is very useful to reexpress the metric components
in the new coordinates:
\be
\eqalign{&\tilde U=\log(\partial_z\partial_{\bar z}K)=U+\log(C_1
\bar C_1)=\log(C_1\bar C_1K_{u\bar u}),\cr
&\tilde V=\log(\partial_w\partial_{\bar w}K)=V+\log(C_2
\bar C_2)=\log(C_2\bar C_2K_{v\bar v}).\cr}\label{newmetric}
\ee
By the same arguments as before we now conclude that the
theory necessarily possesses at least two commuting $U(1)$
isometries, i.e.
$\Phi=\Phi(z+\bar z,w+\bar w)$,
$\tilde U=\tilde U(z+\bar z,w+\bar w)$ and
$\tilde V=\tilde V(z+\bar z,w+\bar w)$.
It follows that the potential $K$ is, up to K\"ahler transformations,
also a function of only $z+\bar z$ and $w+\bar w$,
$K=K(z+\bar z,w+\bar w)$.
The differential equations for the background functions now take the
following form
\be
\eqalign{&\partial_z(\tilde V-2\Phi)=e^{\tilde U},\cr
&\partial_w(\tilde U-2\Phi)=e^{\tilde V},\cr
&\partial_z^2(\tilde U-2\Phi)=\partial_we^{\tilde U},\cr
&\partial_w^2(\tilde U-2\Phi)=\partial_ze^{\tilde
V}.\cr}\label{twoiso}
\ee
The last two equations, which correspond to
$\beta_{u\bar u}^G=0$ and $\beta_{v\bar v}^G=0$,
can be combined and one finds after some algebra
\be
\partial_w(\tilde V-2\Phi)-\partial_z(\tilde U-2\Phi)={\rm constant}.
\label{combined}
\ee

Let us eliminate the dilaton field $\Phi$ from the above equations
in order to obtain the constraints that the string field equations
impose on the geometry, i.e. on $K$. Specifically differentiate
the first two equations in (\ref{twoiso}) with respect to $w$, $z$
respectively. Then we obtain
\be
\partial_w(\partial_z\tilde V-e^{\tilde U})=
\partial_z(\partial_w\tilde U-e^{\tilde V}).\label{intcond}
\ee
This condition is solved by $\tilde V-\tilde U=K_z-K_w+
f(z+\bar z)+g(w+\bar w)$. Substituting this expression into
(\ref{combined}) we
obtain that $f$ and $g$ are linear.
Thus the background geometry has to obey the following constraint:
\be
K_{ww}e^{K_w+c_{2}(w+\bar w)}=K_{zz}e^{K_z+c_1(z+\bar
z)}.\label{finalcon}
\ee
($c_1,c_2$ are arbitrary constants.)
Now we can also compute the dilaton field:
\be
2\Phi=\log K_{zz}-K_w-c_1(z+\bar z)+{\rm constant}.\label{dilfin}
\ee
Finally, the central charge deficit, eq. (\ref{dckahtor}), becomes
\be
\delta c=-3\alpha'(c_1-c_2).\label{dcfinal}
\ee

So far we have considered the conditions on the
backgrounds which are imposed by $N=2$
supersymmetry. For a background to lead to enhanced $N=4$
supersymmetry one expects that non-renormalization theorems \cite{papa}
are valid. These imply that $\delta c=0$.
When the dilaton is zero the condition for N=4 supersymmetry was
derived in
\cite{GHR} and it states that the quasi-K\"ahler potential has to
satisfy
the flat Laplace equation.
This is precisely case (i), eq. (\ref{i}), after a trivial rescaling
of the
coordinates:
\be
K_{u\bar u}=-K_{v\bar v}.\label{ntor}
\ee
Thus $K$ has to satisfy the four-dimensional Laplace equation
\be
(\partial_u\partial_{\bar u}+\partial_v\partial_{\bar
v})K=0.\label{lapl}
\ee
This is the generalization of the hyper-K\"ahler condition for spaces
with antisymmetric tensor field.\footnote{However eq. (\ref{ntor})
is presumably not necessary (but sufficient) for $N=4$
supersymmetry in the presence of non-trivial dilaton field.}

The dilaton field is simply given as
\be
2\Phi=\log K_{u\bar u}+{\rm constant}.\label{dilnfour}
\ee
This has the important consequence that the four-dimensional
metric in the Einstein frame is flat:
\be
G_{\mu\nu}^{{\rm Einstein}}=e^{-2\Phi}G_{\mu\nu}^\sigma=\delta_{\mu
\nu}.\label{einsteinm}
\ee
Therefore, in the Einstein frame,
the nontrivial backgrounds with N=4 symmetry
are entirely given by non-trivial dilaton and axion field
configurations. In fact,
the solutions
of the dilaton equation (\ref{dilnfour}) have a very close relation
to
the axionic instantons of \cite{rey}.
Specifically from eq. (\ref{dilnfour}) we get for example that
$H_{u\bar u v}=K_{u\bar uv}=2e^{2\Phi}\partial_v\Phi$.
Thus, equation (\ref{dilnfour})
implies the following relation:
\be
{\rm d}\Phi=\pm {1\over 2}e^{-2\Phi}H^*.\label{selfdual}
\ee
This relation is nothing else than the self-duality
condition on the dilaton-axion field. Its solutions are
called axionic instantons. However note that an instanton
configuration which provides a solution of eq. (\ref{selfdual})
does not necessarily solves the Laplace equation (\ref{lapl}).
Therefore not any axionic instanton solution is expected
to be $N=4$ supersymmetric.

The form of the solutions of the Laplace equation depends on the
number of isometries of the theory. For the case with two
translational $U(1)$ Killing symmetries, i.e. $K=K(u+\bar u,v+\bar
v)$
the most general solution of (\ref{lapl}) looks like
\be
K=iT(u+\bar u+i(v+\bar v))-i\bar T(u+\bar u-i(v+\bar v)).\label{f}
\ee
For the case with one traslational isometry the general solution
becomes
\be
K(u+\bar u,v,\bar v)=i\int{\rm d}\beta T(\beta,v+\beta(u+\bar u)
-\beta^2\bar v)+{\rm c.c.}\label{g}
\ee
where in both (\ref{f}), (\ref{g}) T is an otherwise arbitrary
function.

\subsection{Dual, K\"ahlerian space which are Ricci flat}
\setcounter{equation}{0}

Let us now
construct the dual spaces for the solutions
of the Laplace equation with one or two isometries,
(\ref{f},\ref{g}).
We will perform a duality transformation  on the chiral $U$-field
replacing it by a twisted chiral field $\Psi$. The Legendre
transformed potential $\tilde K$ will only contain twisted
fields and will be therefore a true K\"ahler function
leading to a non-compact K\"ahler space without torsion.

Doing the Legendre transform we obtain the following line element
\be
{\rm d}s^2={1\over K_{uu}}({\rm d}z-K_{uv}{\rm d}v)({\rm d}{\bar
z}-K_{u\bar
v}{\rm d}{\bar v})-K_{v\bar v}{\rm d}v{\rm d}{\bar v}
\label{riflat}
\ee
where $K(u+\bar u ,v,\bar v)$ is the original quasi-K\"ahler
potential that satisfies the Laplace equation $K_{uu}+K_{v\bar v}=0$
and $z,\bar z$ are the dual coordinates defined via the Legendre
transform $z+\bar z =K_{u}$.
The coordinates $v,\bar v, z,\bar z$ are now the K\"ahler
coordinates.
The Laplace equation implies that the determinant of the K\"ahler
metric
(\ref{riflat}) is constant so we obtain a Ricci flat K\"ahler
manifold.
The dual dilaton is consequently constant.

The general solution to the 4-d Laplace equation with one isometry
can be
written as in (\ref{g}).
Let us introduce the notation
\be
<T>\equiv \int{\rm d}\beta T(\beta,v+\beta(u+\bar u)-\beta^2\bar v).
\label{nota}
\ee
and the function
\be
Z(u+\bar u,v,\bar v)=K_{u}=i<\beta(T_{v}-\bar T_{\bar v})>
\label{def}
\ee
and we should remember that $z+\bar z=Z(u+\bar u,v,\bar v)$.
Then the line element (\ref{riflat}) can be written in the form
\be
{\rm d}s^2={1\over G}({\rm d}z-A_{v}{\rm d}v)({\rm d}{\bar z}-\bar
A_{\bar
v}{\rm d}{\bar v})+G{\rm d}v{\rm d}{\bar v}
\label{bun}
\ee
where,
\be
G={\partial Z\over \partial u}\;\;,\;\;A_{v}={\partial Z\over
\partial
v}\;\;,\;\;{\bar A}_{\bar v}={\partial Z\over \partial \bar v}
\label{def1}
\ee
The interpretation of the metric (\ref{bun}) is as follows:
The $G{\rm d}v{\rm d}{\bar v}$ part describes the metric of a 2-d
Riemann
surface (generically non-compact). The metric depends also on $z+\bar
z$.
For fixed $z+\bar z$, $A_{v},{\bar A}_{\bar v}$ describe a flat line
bundle on the Riemann surface.
The metric (\ref{bun}) is that of a flat complex line bundle on the
Riemann
surface.
The functions $G$, $A_{v},{\bar A}_{\bar v}$ are harmonic.

The metric (\ref{bun}) describes a large
class of 4-d non-compact Calabi-Yau manifolds, which are also
hyper-K\"ahler.
The associated $\s$-models have N=4 superconformal symmetry and $c=6$
($\tilde
c =2$).
The manifolds have generically asymptotically flat regions as well
as curvature singularities.

Let us briefly display a simple example choosing
\be
T=-i\gamma(\beta)e^{u+\bar u+{v\over\beta}-\beta\bar v}.\label{exg}
\ee
Then the potential becomes
\be
K=e^{u+\bar u}\phi(v,\bar v),\qquad \phi(v,\bar v)=
\int{\rm d}\beta\gamma(\beta)\lbrack e^{{v\over\beta}-\beta\bar v}+
e^{{\bar v\over\beta}-\beta v}\rbrack .\label{exga}
\ee
In turn, the dual space is determined by the following
K\"ahler potential:
\be
\tilde K=(z+\bar z)\log(\psi+\bar\psi)-(\psi+\bar\psi)\log
\phi(v,\bar v).\label{dualexg}
\ee
The intergral in (\ref{exga}) can be explicitly performed if we
choose
$\gamma(\beta)=e^{-A\over\beta}\beta^{\nu-1}$:
\be
\phi(v,\bar v)={\rm constant} \biggl(\sqrt{{A-v\over\bar
v}}\biggr)^\nu
K_\nu(2\sqrt{(A-v)\bar v})+{\rm h.c.}\label{integral}
\ee
Here $K_\nu$ is the Bessel function with complex argument.

Let us study now the (more symmetric) special case of (\ref{bun})
with two isometries, i.e. $K(u+\bar u,v+\bar v)$.
If we paramertrize, $u=r_{1}+i\theta$, $v=r_{2}+i\phi$ then K is of
the form
$K(r_{1},r_{2})=iT(r_{1}+ir_{2})-i\bar T(r_{1}-ir_{2})$.
Introducing a new complex coordinate $z=r_{1}+ir_{2}$, we can write
the metric
(\ref{bun}) in the following suggestive form
\be
{\rm d}s^{2}={Im{\rm T}\over 2}{\rm d}z{\rm d}{\bar z}+{2\over Im{\rm
T}}
({\rm d}\theta +{\rm T}{\rm d}\phi)({\rm d}\theta +{\bar {\rm T}}{\rm
d}\phi)\label{tor}
\ee
where T$(z)$ is an arbitrary meromorphic function.
It is crucial to note that the metric (\ref{tor})
is $not$ written in K\"ahler coordinates.
Such coordinates are $v,\bar v$ and $w,\bar w$ with $w+\bar
w=iT'(r_{1}+ir_{2})-i\bar T'(r_{1}-ir_{2})$ and $w-\bar w =2i\theta$.

Now the interpretation of the metric (\ref{tor}) is straightforward:
If we take $\theta,\phi$ to be angular variables, then they
parametrize
a 2-d torus, with modulus T$(z)$ which depends holomorphically
on the rest of the coordinates and conformal factor proportional
to $1/Im $T.
The zeros and poles of the Riemann tensor are determined by the zeros
and
poles (or essential singularities) of the function T$(z)$.

This solution (with a different interpretation) was found in
\cite{NCCY},
where some global issues were also addressed\footnote
{Some generalizations of this idea to more dimensions were recently
presented
in \cite{hu}.}.
We should note that as in \cite{NCCY} a full invariance under the
torus
modular group, $T\rightarrow T+1$ and $T\rightarrow -1/T$ can be
implemented
by a holomorphic coordinate transformation in $z$, which will modify
$Im$T to a modular invariant in the first part of (\ref{tor}).
It was also argued that such a metric might receive higher order
corrections.
However we have just shown that this metric is the dual of the family
of
wormhole solutions which are absolutely stable as CFTs due to their
N=4
superconformal symmetry and it does possess a hyper-K\"ahler
structure
although not easily visible in this coordinate system.

The 4-d non-compact CY manifolds presented in this section constitute
a
large class of exact solutions to superstring theory with extended
supersymmetry.
A detailed analysis of their structure as well as their Minkowski
continuations is beyond the scope of this work and is reserved for
future
study.

\renewcommand{\theequation}{\thesection.\arabic{equation}}
\setcounter{equation}{0}

\section{S Duality and the Axion}

In 4-d we can trade the antisymmetric tensor $B_{\mu\nu}$ for a
scalar field
$b$, the axion.
The $\beta$-function equation for $B_{\mu\nu}$ imply that we can
write
\be
\partial_{\mu}b={{\varepsilon_{\mu}}^{\nu\rho\sigma}\over 6\sqrt{{\rm
det}G^{Einst}}}e^{-4\Phi}H_{\nu\rho\sigma}\label{bdef}
\ee
where
\be
G^{Einst}_{\mu\nu}=e^{-2\Phi}G^{\s}_{\mu\nu}\label{phy}
\ee
and $G^{\s}$ is the $\s$-model metric.
Then the Bianchi identity, $dH=0$ becomes the equation of motion
of the axion field
\be
\nabla^{\mu}\left(e^{4\Phi}\nabla_{\mu}b\right)=0
\ee
If we define\footnote{Since we are in Euclidean space $s_{\pm}$ are
light-cone like variables. They become complex conjugates in
Minkowski space.}
\be
s_{\pm}=\pm b+e^{-2\Phi}
\ee
then the 1-loop effective action which gives the $\beta$-function
equations as
equations of motion is
\be
S_{eff}=\int\sqrt{{\rm det}G}\left(
R+2G^{\mu\nu}{\d_{\mu}s_{+}\d_{\nu}s_{-}\over
(s_{+}+s_{-})^2}+{2\delta c\over
3}{1\over (s_{+}+s_{-})}\right)\label{eff}
\ee
written in the Einstein frame (\ref{phy}).
If $\delta c =0$, then $S_{eff}$ is invariant under SL(2,R)
transformations:
\be
s_{+}\rightarrow {as_{+}+b\over cs_{+}+d}\;\;,\;\;s_{-}\rightarrow
{-as_{-}+b\over cs_{-}-d}\;\;,\;\; ad-bc=1\label{sl2}
\ee
At higher orders the $SL(2,R)$ symmetry is broken down to $SL(2,Z)$
known as
the S-duality symmetry.

An important question is: does S-duality preserves spacetime
supersymmetries?
To answer the question in the affirmative, we need to observe that
spacetime fermions transform homogeneously, as forms, under
S-duality.
In particular, S-duality transformations of the Dirac operators
(whose zero
mode spectrum determine the
number of unbroken supersymmetries) correspond to gauge
transformations.
Thus, they do not affect the zero mode spectrum and consequently the
number
of unbroken supersymmetries.
If the worldsheet theory has N=4 superconformal symmetry then the
1-loop
equations have no higher order corrections.
Then S-duality generates exact N=4 backgrounds.

We will study separately the action of S-duality on solutions to the
$\beta$-function equations corresponding to cases (i,ii,iii) in
section
4.1.

(i) The N=4 solutions found in section 4.1 translate as follows in
our new
variables,
\be
s_{+}=C\;\;,\;\; s_{-}=2e^{-2\Phi}-C
\ee
where
\be
\square e^{2\Phi}=0
\ee
$C$ is a constant and the metric is flat.
Under an SL(2,Z) transformation we will stay in the same class
of solutions with $C'=(aC+b)/(cC+d)$ and $e^{2\Phi'}=(cC+d)^2
e^{2\Phi}-2c(cC+d)$.

S-duality is a symmetry of the string effective action in 4-d
and its status is obscure although some progress has been made
towards
elucidating its implications for string theory (see for example
\cite{S}
and references therein). In particular a worldsheet understanding of
the
symmetry is lacking.
It is our opinion that class (i) of string solutions is a very useful
ground to study the implications of S-duality, due to their high
symmetry (N=4) and the fact that they are exact solutions to string
theory.

(ii) In this case the quasi-K\"ahler potential satisfies
(\ref{solveb}).
The solutions found in this class are the N=2 preserving
O(2,2,R) transforms
of the K\"ahler solutions in section 3.
In order to have $\delta c =0$ we must set the constant $c_{1}$
in (\ref{solveb}) to zero
(If at higher orders, $\delta c \not= 0$ then $S$-duality
will not produce new solutions).
The Einstein metric here is not flat
\be
G^{Einst}=\left(\matrix{ 0&e^{K_{\rm v}}&0&0\cr
e^{K_{\rm v}}&0&0&0\cr
0&0&0&1\cr 0&0&1&0\cr}\right)
\ee
and
\be
s_{+}=-{\rm v+\bar v}\;\;,\;\;s_{-}=-{2\over K_{\rm vv}}+{\rm v+\bar
v}
\label{sss}
\ee
Acting on (\ref{sss}) with SL(2,Z) transformations generates new
solutions.

In case (iii) the fields $s_{\pm}$ cannot be written explicitly as
functions
of the K\"ahler potential, so we can not proceed further.

For K\"ahlerian backgrounds (H=0) the only S-duality transformation
which stays
in this class, changes the sigh of the dilaton (inverts the string
coupling).

Another set of new solutions can be obtained from the tensor
product theory ${SU(2)_{k}\over U(1)}\times{SL(2,R)_{k+4}\over U(1)}$
\cite{C2} by applying S-duality, but we will not delve further into
this.

\section{Conclusions}
\setcounter{equation}{0}

We have examined some four-dimensional superconformal theories with
N=2 and N=4
superconformal symmetry (classical solutions to superstring theory).
We show that there exists a plethora of such theories with
non-trivial
metric, dilaton and antisymmetric tensor field.

Our solutions are classified in two classes:

(i) Those that are based on a K\"ahler manifold (when
$H_{\mu\nu\rho}=0$).

(ii) The non-K\"ahlerian solutions with non-zero torsion.

These two subclasses are related by $Z_{2}$ duality transformations
(when
isometries are present).
$Z_{2}$ duality interchanges the roles of untwisted and twisted
chiral
superfields and act in a manifest N=2 preserving fashion.
There is a generic subgroup of $O(d,d,Z)$ duality and a corresponding
subgroup of its conformal
deformation
partner,
$O(d,d,R)$, which are shown to preserve the extended superconformal
invariance
(N=2,4).
They are used to obtain a more general class of solutions where
a superfield formulation is not manifest.
It is an interesting problem to formulate the N=2 preserving O(d,d)
action
in a manifestly N=2 supersymmetric form.

Another way of enlarging the class of solutions is via S-duality,
which is
a symmetry when $\delta c =0$, as it happens in the presence of N=4
symmetry.

In the K\"ahlerian case we show that the presence of a non-trivial
dilaton field implies the presence of an isometry in the background
data
(K\"ahler metric and dilaton).
The current associated to this isometry is the bosonic part of the
N=2
U(1) current (its presence is due to the non-trivial dilaton).
A non-trivial dilaton implies also a different LG hypersurface
constraint
which generalizes the vanishing of the superpotential in the constant
dilaton
case.
It is interesting that we obtain these modifications from the
requirement
of consistently coupling the $\sigma$-model to N=2 supergravity.
In particular the constraints above come directly from the scalar
constraint
associated to the auxiliary field of the N=2 supergravity multiplet.
Among the K\"ahlerian solutions we find a large class of
(non-compact)
Ricci-flat (CY) manifolds with one isometry.
This class of solutions generalizes the compact 4-d Ricci flat
manifolds (K3).
A special case of the solutions above (with two isometries) is that
of ref.
\cite{NCCY} found in a slightly different context.
These CY manifolds are duals of non-zero torsion solutions with N=4
symmetry.

The N=2,4 solutions with non-zero torsion are classified in terms of
zero
one and two isometries in the special (quasi-K\"ahler) coordinate
system.
We have derived also their K\"ahlerian duals.

The Hyper-K\"ahler solutions we find have non-trivial metric
in the Einstein frame.
On the contrary, the N=4 solutions with non-zero torsion have always
a flat
Einstein metric and some of them (those with two isometries) give all
the
possible supersymmetric axion-dilaton instantons.

It is also observed that, contrary to expectations, isometries are
generic
in 4-d supersymmetric string backgrounds.

Upon analytic continuation of the Euclidean solutions we have found
we can obtain many cosmological solutions to superstring theory whose
spacetime
properties deserve further study.

\vskip .5cm
\noindent
{\bf Acknowledgments} \\

We would like to thank L. Alvarez-Gaum\'e and S. Ferrara for
discussions.
This work was partially supported by EEC grants, SC1$^{*}$-03914C
and SC1$^*$-CT92-0789.
\noindent

\newpage
\setcounter{section}{0}
\setcounter{equation}{0}

\renewcommand{\thesection}{Appendix \Alph{section}.}
\renewcommand{\theequation}{A.\arabic{equation}}
\section{}
\vskip .8cm
\centerline{\bf Scalar Constraint in N=2 d=2 Supergravity}
\centerline{\bf and its equivalence to Generalized N=2 L-G
 Equations.}
\vskip .5cm

In this appendix we will show that the (2,0) and (0,2)
$\beta$-function
equations (\ref{ric},\ref{phikah}) of an $N=2$ superconformal
system, follows from a classical scalar constraint due to the
auxiliary field of
the $N=2$
supergravity multiplet, in the presence of a non trivial dilaton
background
${\Phi} R^{(2)}$ and non zero chiral deformation $\int W(\phi_i)$.
Also we will
show how  the non trivial dilaton background implies the existence of
an
isometry in the metric, and how the $N=2$ LG equation ($
w(\phi_i)=0$) are
generalized in the presence of non trivial dilaton\footnote{In the
case of
(p,0) worldsheet supersymmetry without dilaton the constraints for
the
superpotential were recetly analysed in \cite{hpt}.}.

First we would like to extend the ${\Phi} R^{(2)}$ dilaton background
to a
generalized
$N=2$ one, where the other components of the gravitational
supermultiplet explicitly appear.
We will do this extension in the superconformal gauge where the
gravitational
multiplet can be  represented  by a chiral superfield $\hat{\sigma}$
\cite{sup}
\be
\hat{\sigma}=\sigma+\theta_{+} \eta_{+}+\theta_{-}
\eta_{-}+\theta_{+}\theta_{-} H\label{a1}
\ee
and its complex conjugate.

The metric $g_{\alpha\beta}$, the U(1) gauge field $A_{\alpha}$, the
two
gravitini $\chi^{\alpha}_{+}$ $\chi^{\alpha}_{-}$ and the
supergravity
auxiliary field can be expressed in terms of $\sigma$, $\eta_{+}$,
$\eta_{-}$
and H as follows,
\be
g_{\alpha\beta}=\eta_{\alpha\beta}e^{\s+\bar \s}\label{a2}
\ee
\be
A_{\alpha}=i\partial_{\alpha}(\sigma-\bar{\sigma})\label{a3}
\ee
\be
\chi^{\alpha}_{L}=\gamma^{\alpha}\eta_{+}\label{a4}
\ee
\be
\chi^{\alpha}_{R}=\gamma^{\alpha}\eta_{-}\label{a5}
\ee

We assume n chiral superfields
\be
\hat{\phi_{i}}=\phi_{i}+\theta_{+}\psi^{+}_{i}+\theta_{-}\psi^{-}_{i}
+\theta_{+}\theta_{-} F_{i}\label{a6}
\ee
which we want to couple with the supergravity multiplet
$\hat{\sigma}$.

 The most general coupling among the chiral fields $\hat{\phi_{i}}$
and
$\hat{\sigma}$ is given in terms of a general K\"ahler function
$G(\phi_{i}, \bar{\phi_{i}}; \sigma,$$\bar{\sigma)}$ and a general
superpotential
$W(\phi_{i}; \sigma)$ as usual (see section 2.1). In our case both
the
functional form of $G$ and
$W$ are  restricted and this is because the $\hat{\sigma}$ field  has
to define
correctly the background metric and the background gauge field of the
supergravity multiplet as in the above equations; e.g.
$e^{(\sigma+\bar{\sigma})}$ is the conformal factor of the metric
(\ref{a2}).
Once we take that in to account, $G$ and $W$ are given as
\be
G(\phi_{i}, \bar{\phi_{i}}; \sigma,\bar{\sigma})=K(\phi_{i},
\bar{\phi_{i}})+\sigma\bar{\Lambda}(\phi_{i}, \bar{\phi_{i}})+
\bar{\sigma}\Lambda(\phi_{i}, \bar{\phi_{i}})\label{a7}
\ee
and
\be
W(\phi_{i}; \sigma)=w(\phi_{i})e^{\sigma}\label{a8}
\ee
The terms linear in $\sigma$ and $\bar{\sigma}$ in $G$ will give rise
to the
generalization of the dilaton background (see below), while the
K\"ahler
potential $K$ defines the kinetic terms of the $\phi_{i}$
superfields. Notice
also that the $\sigma$ dependence of $W$ in the above equation is
fixed
uniquely due to the $N=2$ U(1) charge conservation.

We are now in a position to examine the algebraic equations which
follows after
the elimination of the auxiliary fields $H$ and $F_{i}$. Setting the
fermions
to zero the relevant part of the $N=2$ lagragian reads
\be
\Delta {\cal L}= K_{i\bar{j}}F_{i}\bar{F_{\bar{j}}}+H
\bar{\Lambda}_{\bar{j}}
\bar{F}_{\bar{j}}+\bar{H}\Lambda_{i}
F_{i}+W_{i}F_{i}+\bar{W}_{\bar{j}}\bar{F}_{\bar{j}}+WH+\bar{W}
\bar{H}\label{a9}
\ee

The equation of motion of the auxiliary field $H$ implies the
constrain
\be
\bar{\Lambda}_{\bar{j}}\bar{F}_{\bar{j}} +W=0 \,\,\,\,\,\,\,\,\,\,
{\rm and}
\,\,\,\,\, H=0.\label{a10}
\ee
The second equation is similar to the the Gauss law.

The $F_{i}$ equations are
\be
K_{i\bar{j}}\bar{F_{\bar{j}}}+\bar{H}\Lambda_{i}+W_{i}=0\label{a11}
\ee
Combining the above equations we obtain the following important
relation among
$W$, $W_{i}$ $K_{i\bar{j}}$ and $\bar{\Lambda}_{\bar{j}}$
\be
\bar{\Lambda}_{\bar{j}}[K_{i \bar{j}}]^{-1}W_{i}=W\qquad {\rm
and}\qquad {\rm
cc}\label{a12}
\ee

Let us define by $\xi_{i}$ the holomorphic vector
\be
\xi_{i}=\bar{\Lambda}_{\bar{j}}[K_{i \bar{j}}]^{-1}\label{a13}
\ee
then, the above relation implies the following symmetry for $W$
\be
W(\phi_{i}-\epsilon \xi_{i}; \sigma + \epsilon)=0\label{a14}
\ee
Indeed for $\epsilon \rightarrow 0$ we have $\epsilon
\xi_{i}W_{i}=\epsilon W$.
For $\epsilon =i|\epsilon|$ the above symmetry impose the charge
conservation
of the superpotential. Observe that this symmetry makes sense for any
non
trivial $W$ only if $\xi_{i}$ is an holomorphic function in terms of
the fields
$\phi_{i}$, which implies that
\be
\bar{\Lambda}_{\bar{j}}=K_{i\bar{j}}\xi_{i}(\phi_{i})\label{a15}
\ee
and after anti-analytic integration over $\bar{\phi}_{\bar{j}}$ we
have
\be
\bar{\Lambda}=K_{i}\xi_{i}(\phi_{i})+\rho(\phi_{i})\label{a16}
\ee
and similarly
\be
\Lambda=K_{\bar j}{\bar \xi}_{\bar j}(\bar \phi_{\bar j})+{\bar
\rho}\bar
(\phi_{\bar j})\label{a166}
\ee
The above equations drastically constrain the K\"ahler metric
$K_{i\bar{j}}$
and the dilaton function $\Phi$ as we will see below.

The dilaton function $\Phi$ is given in terms of $\Lambda $ and
$\bar{\Lambda}$
and can be determined from the $ d\bar{\sigma} d\phi_{i}$ and $
d\sigma
d\bar{\phi}_{\bar{i}}$ kinetic terms,
\be
J^{D}d(\sigma+ \bar{\sigma})+iJ^{A}d(\sigma- \bar{\sigma})=
2(\Lambda_{i}d\phi_{i}d\bar{\sigma}+
\bar{\Lambda}_{\bar{j}}d\bar{\phi}_{\bar{j}}d\sigma)\label{a17}
\ee
where $J^{D}$ and $J^{A}$ are the dilatation current and the bosonic
part of
the $N=2$ U(1) R-current respectively.
\be
J^{D}=\Lambda_{i}d\phi_{i}+
\bar{\Lambda}_{\bar{j}}d\bar{\phi}_{\bar{j}}\label{a18}
\ee
\be
iJ^{A}=\Lambda_{i}d\phi_{i}-
\bar{\Lambda}_{\bar{j}}d\bar{\phi}_{\bar{j}}\label{a19}
\ee

The existence of a dilaton function $\Phi$ requests the integrability
of the
dilatation current,
\be
J^{D}={\rm d}\Phi\label{a20}
\ee
since only then
\be
\int J^{D}d(\sigma+ \bar{\sigma}) =-\int {\Phi} \,\, \square(\sigma+
\bar{\sigma})
=\int {\Phi}\,\, R^{(2)}\label{a21}
\ee
The integrability constrain $J^{D}=d \Phi$ implies the following form
for
$\Phi$
\be
{\Phi} = L(\phi_{i},\bar{\phi}_{\bar{j}})+f(\phi_{i})+
\bar{f}(\bar{\phi}_{\bar{j}}) \label{a22}
\ee
with
\be
L(\phi_{i},\bar{\phi}_{\bar{j}})=K_{i}\xi_{i}=
K_{\bar{j}}\bar{\xi}_{\bar{j}}\label{a24}
\ee

Equations (\ref{a22},\ref{a24}) imply that
$\nabla_{i}\partial_{j}\Phi =0$
and its complex conjugate, which, taking into account that for a
K\"ahler
manifold $R_{ij}=0$, are the (2,0) $\beta_{ij}=0$ and (0,2)
$\beta$-function
equations, (\ref{ric}).
Thus, we see that the consistent coupling of an N=2 $\s$-model to N=2
supergravity implies at the classical level the (2,0) and (0,2)
$\beta$-function equations, which as shown in section 3.1
imply in turn the existence of a killing symmetry in the K\"ahler
potential.
The interpretation of this is as follows.
If the dilaton is constant, the N=2 U(1) current is purely fermionic
and conserved at the classical level.
When the dilaton is non-trivial, then the N=2 U(1) current acquires a
bosonic
contribution ($J^{A}$). In order for it to continue to be conserved,
a killing
symmetry of the K\"ahler manifold is necessary.

As we have explained above (\ref{a14}) implies the classical scalar
constraint
\be
w-\xi_{i}w_{i}=0\label{a25}
\ee
which generalizes the LG hypersurface equation in the presence of non
trivial
dilaton.
This can be easily solved since upon a holomorphic coordinate
transformation
we can rotate the vector $\xi_{i}$ to a specific direction (call it
$\phi_{0}$)
and make its value one.
Then
\be
w(\phi_{0},\phi_{i})=w(\phi_{i})e^{\phi_{0}}
\ee

\def\PL{Phys. Lett. }
\def\NP{Nucl. Phys. }
\def\PR{Phys. Rev. }

\end{document}